\documentclass[useAMS,usenatbib,twocolumn]{mn2e}

\usepackage{amsmath,amssymb}
\usepackage{graphicx,xcolor}
\usepackage{natbib}
\usepackage{url}

\newcommand{\spose}[1]{\hbox to 0pt{#1\hss}}

\newcommand{\erf}{\mathop{\rm erf}\nolimits}

\newcommand{\EXP}{{\sc{exp}}}
\newcommand{\GADGET}{{\sc{gadget-2}}}

\newcommand{\msun}{{\rm\,M_\odot}}

\mathchardef\star="313F

\newcommand{\aj}{AJ}         % Astronomical Journal
\newcommand{\apj}{ApJ}       % Astrophysical Journal
\newcommand{\mnras}{MNRAS}   % Monthly Notices
\newcommand{\aap}{A\&A}      % Astronomy and Astrophysics
\newcommand{\apss}{Ap\&SS}   % Astrophysics and Space Science
\title[\(l=1\) instabilities]{New dipole instabilities in spherical
  stellar systems}

\author[M.~D.~Weinberg]{Martin D. Weinberg\thanks{E-mail:
    mdw@umass.edu}\\
  Department of Astronomy\\ University of Massachusetts, Amherst
  MA 01003-9305, USA }

\begin{document}

\label{firstpage}

\pagerange{\pageref{firstpage}--\pageref{lastpage}} \pubyear{2023}

\maketitle

\begin{abstract}
  Spherical stellar systems have weakly-damped response modes. The
  dipole modes are seiche modes. The quadrupole are zero pattern-speed
  prolate modes, the stable precursors to the radial-orbit instability
  (ROI). We demonstrate that small wiggles in the distribution
  function (DF) can destabilise the dipole modes and describe the
  newly identified instabilities in NFW-like dark-matter (DM) halos
  and other power-law spherical systems.  The modes were identified in
  N-body simulations using multivariate singular spectrum analysis
  (MSSA) and corroborated using linear-response theory.  The new mode
  peaks inside the half-mass radius but has a pattern speed typical of
  an outer-halo orbit.  As it grows, the radial angle of the eccentric
  orbits that make up the mode correlate and lose angular momentum by
  a resonant couple to outer-halo orbits.  This leads to an unsteady
  pattern with a density enhancement that swings from one side of the
  halo to another along a diameter, like the orbits that comprise the
  instability. In this way, the dipole mode is similar to the ROI.
  Since the DF found in Nature is unlikely to be smooth and isotropic
  with \(df(E)/dE<0\) necessary for Antonov stability, these modes may
  be ubiquitous albeit slowly growing. Halos that are less extended
  than NFW, such as the Hernquist model, tend to be stable to this
  dipole instability.  We present the critical stability exponents for
  one- and two-power models.  These different critical outer power-law
  exponents illustrate that the gravitational coupling between the
  inner and outer DM halo depends on the global shape of density
  profile.
\end{abstract}


\section{Introduction}
\label{sec:intro}

\subsection{Instabilities in spherical stellar systems}

The stability of spherical stellar systems are the best well-studied
of all astronomical equilibrium models.  The modern contributions
began with work of \cite{Antonov:62,Antonov:73}.  Since then, there
have been three main approaches: the study of global functionals of
energy following Antonov's approach, the analytic derivation of
discrete modes, and numerical simulation.  The early work on the modal
approach is summarised in the monographs by
\citet{Fridman.Polyachenko:84} and followed up by
\cite{Palmer.Papaloizou:87,Weinberg:91c,Palmer:94}.  In brief, one
simultaneously linearises and solves the collisionless Boltzmann
equation and the Poisson equation.  With a careful choice of
phase-space variables and Fourier decomposition, we obtain a solvable
set of algebraic equations for the frequency spectrum.  Using this
dispersion relation for the eigenfrequencies of the system, an
instability is present if at least one of the eigenfrequencies has a
non-zero positive imaginary part, while stability holds if they are
all real or have negative imaginary parts.  The set of stable models
are candidates for describing long-lived astronomical systems,
galaxies in particular.  See \citet{Binney.Tremaine:2008} for a
comprehensive review.

Beyond stability, environmental interactions and assembly history
disturb equilibria.  Like plasma systems, disturbances to a self
gravitating systems can be described by linear operator on a function
space with infinite degrees of freedom.  Mathematically, such
operators have a more complex spectrum than the familiar normal
(eigen) modes of a finite system \citep[e.g.][]{Riesz.Nagy:2012}.  As
in plasmas, self-gravitating stellar systems have a spectrum of both
{\it continuous} and {\it point modes}
\citep{Ichimaru:73,Ikeuchi.etal:74}, and the response of an initially
equilibrium system to a perturbation is a combination of both parts of
the spectrum\footnote{Mathematically, the spectrum of our response
  operator has a standard decomposition into three parts: (1) a point
  spectrum, consisting of eigenvalues; (2) a continuous spectrum,
  consisting of the scalars that are not eigenvalues but have a dense
  range; and (3) a residual spectrum, consisting of all other scalars
  in the spectrum.  The details of the full decomposition are not
  relevant here.}.  The point modes have distinct shapes reinforced by
their own gravity.  In plasma physics, the modes in the point spectrum
are called \emph{Landau} modes \citep{Landau:46} and those in the
continuous spectrum are called van Kampen modes \citep{Kampen:55}. The
initial response in stellar systems is dominated by the continuous
modes which typically phase mix in several dynamical times.  The most
commonly known point mode is the Jeans' instability in a homogeneous
sea of stars \citep[e.g.][]{Binney.Tremaine:2008}.  See
\citet{Lau.Binney:21} for a discussion of van Kampen modes in the
context of stellar dynamics.

This contribution concerns the long, slow evolution of a dark matter
halo and therefore, the behaviour of point modes is key.  The point
modes continue to exist for stable systems but are damped rather than
growing; i.e. their frequencies have negative imaginary parts. For
example, the Jeans' mode in the homogeneous system becomes a damped
mode with increasing velocity dispersion.  This mode can be excited by
a disturbance and persist over time if the damping time is sufficient
long.  \citet{Weinberg:94} extended this idea to spherical systems,
demonstrating the existence of both \(l=1\) and \(l=2\) damped modes
in two ways. First, a self-consistent solution of the linearised
collisionless Boltzmann equation provides a response operator that
defines damped and growing modes (often called the \emph{dispersion
  relation} by analogy with plasma physics) and, second, by N-body
simulation.

While early work on time-dependence of galaxy structure focused on
stability and disc features, modern suites of numerical N-body
simulations
\citep{Illustris:2019,Auriga:2019,NewHorizon:2021,FIRE:2022} make it
clear that the evolution of galaxies are driven by disturbances from
their assembly at early times and their environment in the current
epoch.  The fluctuations from environmental disturbances such as
satellite encounters or Poisson noise from \(N\)-body distributions
may excite these weakly self-gravitating features.  Calculations for
unstable evolutionary modes in galactic discs have found evidence for
point modes supported in various analytic geometries
\citep[e.g.][]{Fouvry.etal:2015,DeRijcke.etal:2019}.

\subsection{Overall plan}

While we can understand the basic governing principles with detailed
mathematical models from Hamiltonian perturbation theory, the
inter-component and environmental interactions that produce key
features of galaxy morphology, such as bars and spiral arms, are hard
if not impossible to study from dispersion relations alone.  This
paper was motivated by simulations of disc galaxies evolving in dark
matter halos
\citep[e.g.][]{Petersen.etal:2019a,Petersen.etal:2019b,Petersen.etal:2019c}.
In these papers, we performed controlled simulations using our
basis-function expansion (BFE) Poisson solver \citep[\EXP]{EXP:2021}.
The BFEs are designed to spatially best represent the degrees of
freedom that we know to describe the collective nature of the
dynamics.  We noticed that \(l=1\) power in the halo appeared quickly
and persisted over the typically 8--10 Gyr of the simulation.  These
same BFE techniques underpin the numerical computation of the
dispersion relations used in \cite{Weinberg:91c,Weinberg:94}.  We
exploit this correspondence for additional insight.

More recently, \citet{Weinberg.Petersen:2021} combined the BFE
representation of the possibly unknown dynamics in simulations with
the knowledge acquisition algorithm known as \emph{multivariate
  singular spectrum analysis} \citep[mSSA,
e.g.][]{Ghil.Vautard:1991,Golyandina.etal:2001,Ghil.etal:2002}.  This
allows us to find couplings that may be too hard to predict otherwise.
A detailed investigation of the persistent \(l=1\) halo power
revealed the existence of response mode that slowly grows for a
variety of plausibly realistic dark-matter halo conditions.

In short, the plan of the remainder is as follows. We begin with a
description of methods and models in Section \ref{sec:methods}. We
move on to a description of the new instability in Section
\ref{sec:instability}.  Section \ref{sec:analysis} characterises the
instability in detail using the combination of BFE representation and
mSSA.  This helps determine the nature of the mode and its physical
mechanism.  We verify that the unstable mode is not code dependent in
Section \ref{sec:verify} by recovering the results of our \EXP\
simulations using \GADGET.  Section \ref{sec:linear} demonstrates that
the domain of instability found in the N-body simulations is predicted
by the linear perturbation theory and corroborates the empirical
findings of the previous sections.  We end with some final discussion
and a summary in Section \ref{sec:summary}.

\section{Simulations and methods}
\label{sec:methods}

We review the key features of \EXP\ (Sec. \ref{sec:nbody}), mSSA
(Sec. \ref{sec:mssa}), and linear response theory (Sec. \ref{sec:lrt})
followed by a description of the model families used in this
investigation (Sec. \ref{sec:models}).

\subsection{N-body methods}
\label{sec:nbody}

We require a description of the potential and force vector at all
points in physical space and time to compute the time evolution for an
N-body system. We accomplish this using a biorthogonal basis set of
density-potential pairs that solve the Poisson equation.  We generate
density-potential pairs using the basis function expansion (BFE)
algorithm described in \citet{Weinberg:99} and implemented in \EXP. In
the BFE method \citep{Clutton-Brock:72, Clutton-Brock:73,
  Hernquist.Ostriker:92}, a system of biorthogonal potential-density
pairs are calculated and used to approximate the potential and force
fields in the system.  The functions are calculated by numerically
solving the Sturm-Liouville equation for eigenfunctions of the
Laplacian. The full method is described precisely in
\citet{Petersen.etal:2022}.  Appendix \ref{sec:biorth} provides
technical details and a presentation of simulation parameters.

For comparison, we also use \GADGET\ \citep{Springel:2005} which
computes gravitational forces with a hierarchical tree algorithm.  We
only need the tree-gravity part of \GADGET\ to follow the evolution of
a self-gravitating collisionless N-body system. We choose \GADGET\
because it is well-known in the community but any modern gravity-only
Lagrangian code would suffice.  We compute the softening following the
algorithm from \citet{Dehnen:2001}.  In practice, \GADGET\ does not
conserve linear momentum as well as \EXP, and this demands a
recentring computation for inter-comparison.  The density centre for
the snapshots produced by each code is evaluated using kd-tree
nearest-neighbour estimator.  The density and potential fields for the
evolved phase space for each code are then computing using the same
BFE expansion used by \EXP.

\subsection{Multivariate Singular Spectrum Analysis}
\label{sec:mssa}

Singular spectrum analysis \citep[SSA, see][]{Golyandina.etal:2001}
extracts correlated signals from a time series by analysing the
covariance matrix of a series at successive times lags with itself.
For example, consider a pure sinusoidal signal.  When the lag equals
the period, the corresponding cross-covariance element will be large,
otherwise it will tend to vanish by interference.  This method works
for aperiodic signals as well.  For a time series with some arbitrary
but coherently varying signal, its correlated temporal variations will
reinforce when the lag coincides with its natural time scale.  SSA has
similarities with its cousin, PCA (principal component analysis).  In
PCA, the typical covariance matrix consists of a single field variable
observed at multiple positions. In SSA, the covariance matrix consists
of a single time series observed in different time windows; the
resulting components derived through SSA are \emph{temporal} principal
components.

The BFE coefficients for an N-body simulation at any point in time
(see Appendix \ref{sec:biorth}) are a spatial analysis of the
gravitational field. The full simulation is represented by multiple
time series, one for each coefficient. The entire BFE in time encodes
a time-varying interrelated spatial structure.  To apply SSA to this
situation, we extend the analysis of a single time series to multiple
time series simultaneously; this is mSSA and the application to
dynamics is described in
\citet{Weinberg.Petersen:2021,Johnson.etal:2022}.  In short, mSSA
applied to a BFE in time provides a combined spatial and temporal
spectral analysis.  The dominant eigenvectors reveal key correlated
dynamical signals in our simulation.  The analyses of simulations in
Section \ref{sec:instability} rely on mSSA for isolating correlated
dynamical signals.

\subsection{Linear response theory}
\label{sec:lrt}

We use the \emph{matrix-method} solution of the linearised
collisionless Boltzmann equation to estimate the location of the point
modes to guide our interpretation of the N-body simulations.  This
semi-numerical method has been used by the author and many others
(op. cit., Sec. \ref{sec:intro}) to explore secular evolution.  In
short, this method discretises the variation in the density and
potential fields under a Hamiltonian flow by expanding them in
biorthogonal potential--density functions that mutually solve the
Poisson equation.  With this \emph{built-in} solution to the Poisson
equation, the response of the Hamiltonian system to an applied
perturbation may be derived by splitting the slower temporal evolution
driven by the perturbation from the faster characteristic orbital
times, and then averaging over these latter degrees of freedom.  This
method can be used to estimate the evolution of galactic system to a
wide variety of distortions such as minor mergers and predict the
features of disc and halo interactions.  Here, we are interested the
answer to a very specific question: under what condition is the
\emph{output} response of the system the same as the \emph{input}
perturbation?  This is a generalised eigenvalue problem and the
solutions are the \emph{point modes} described earlier.  The complex
frequency \(\omega\) that satisfies this condition determines whether
this response grows (\(\Re(\omega)>0\)) or decays (\(\Re(\omega)<0\)).
More details on computing this eigenvalue problem and finding the
shape of the shape of the mode can be found in Appendix
\ref{sec:disper}.

\subsection{Models}
\label{sec:models}

Nearly all of idealised dark-matter halo models have infinite extent
and some have infinite mass.  Their isotropic phase-space distribution
functions, \(f(E)\), have \(df/dE<0\) everywhere and are stable by
Antonov's theorems (op. cit.).  This idealisation does not describe
the true structure of a halo, however.  By their nature, halos will
have varying anisotropy, dark-matter density floors
\citep{Diemer.Kravtsov:2014} and presumably, `bumps' and `wiggles' in
their distribution owing to incomplete mixing from assembly and
mergers.  We will demonstrate that these types of deviations at
specific scales drive growth of otherwise damped modes.

Our fiducial model is NFW-like \citep{Navarro.Frenk.ea:97} with
\(c=10\) and an outer truncation to provide finite mass.  Although
this is a smaller concentration than a cosmological dark-matter halo
with the Milky Way mass, this concentration provides a good match to
the Milky Way rotation curve when combined with a realistic Milky Way
disc \citep{McMillan:2017,Petersen:pc1}.  The NFW halo is slightly
modified at small radii and large radii, to account for dissipation
and evolution at small radii and provide a finite mass at large radii.
We adopt:
\begin{equation}
  \rho(r) \propto \frac{1}{r+r_c}\frac{1}{(r + r_a)^2} {\cal T}(r;
  r_t, \sigma_t)
  \label{eq:NFW}
\end{equation}
where \(r_c\) is the inner \emph{core} radius, \(r_a\) is the
characteristic scale radius, \(r_t\) is the outer truncation radius,
and \(\sigma_t\) is the truncation width.  The truncation function,
\({\cal T}\), is unity at radii \(r\ll r_t\) and tapers the density to
zero beyond \(r_t\) with a width \(\sigma_t\) as follows:
\begin{equation}
  {\cal T}(r; r_t, \sigma_t) = \frac12\left[1 -
    \erf\left(\frac{r - r_t}{\sigma_t}\right)\right].
  \label{eq:trunc}
\end{equation}
Here, the inner core radius is a convenience that obviates computing
very high frequency orbits for a negligible measure of phase space
when performing numerical dispersion relations from linear theory
(Sec. \ref{sec:linear} and App. \ref{sec:disper}).  We choose \(r_c\)
smaller than any scale of interest and, therefore, the value of
\(r_c\) has very little effect on the N-body simulations or the
dispersion relation.  We also study models from the more general
two-power form inspired by the NFW model:
\begin{equation}
  \rho(r) \propto \frac{1}{(r+r_c)^\alpha}\frac{1}{(r + r_a)^\beta}
  {\cal T}(r; r_t, \sigma_t)
  \label{eq:two-power}
\end{equation}
For dark-matter-inspired models, we assign \(\alpha=1\) and
\(\beta\in[2-4]\).  The \citet{Hernquist:90} model, for example, has
\(\alpha=1, \beta=3, r_t=\infty\).  We also consider truncated single
power-law models with \(\alpha\in[1-3]\) and \(\beta=0\).

We adopt \emph{virial} units with unit mass and unit radius
corresponding to the virial mass and virial radius, respectively, with
\(G=1\).  In these units, our fiducial model has the parameters:
\(r_c=10^{-5}, r_a=10^{-1}, r_t=1.2, \sigma_t=0.1\).  The fiducial
simulations use \(N=2\times10^7\) particles.  Some auxiliary tests use
\(N=10^6, 10^7, 10^8\).  The main results are robust to the choice of
\(N\) in this range.  Larger particle numbers better discern
phase-space details.

For each model, we construct a phase-space distribution function in
energy \(E\) and angular momentum \(L\) using the Osipkov-Merritt
generalisation of the Eddington inversion
\citep{Osipkov:79,Merritt:85,Binney.Tremaine:2008}.  We use a negative
energy convention; the energies at the centre of the model are
smallest (most bound) and are largest at the truncation radius (least
bound).  This generalisation assumes a univariate distribution
function \(f(Q) = f(E + L^2/2r_q^2)\) where \(L\) is the total angular
momentum and \(r_q\) is the anisotropy radius.  For
\(r_q\rightarrow\infty, f(Q)\rightarrow f(E)\).  We do not consider
the more limited circularly anisotropic case,
\(f(Q) = f(E - L^2/2r_q^2)\).  In many cases, we use the pure
Eddington method for an isotropic distribution function \(f=f(E)\).
We realise a phase-space point by first selecting random variates in
energy \(E\) and angular momentum \(L\) by the acceptance-rejection
technique from the distribution \(f(Q)\). Then, the orientation of the
orbital plane and the radial phase of the orbit are chosen by uniform
variates from their respective ranges. This determines position and
velocity. Particle masses are identical. The inversion is not
guaranteed to provide physical solutions, and we restrict the models
to parameters values that yield positive mass densities.  We explore a
suite of varying \(r_q\) for the NFW-like model described by equation
(\ref{eq:NFW}).

\section{Key findings}
\label{sec:instability}

\begin{figure}
  \includegraphics[width=0.45\textwidth]{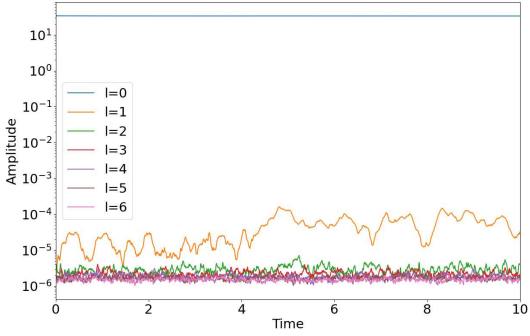}
  \caption{The amplitude for harmonic orders \(l=0,\ldots,4\).  This
    amplitude, which is the square root of the gravitational power, is
    plotted for an easy comparison with the background.  The \(l=0\)
    amplitude contains the equilibrium field and is nearly constant in
    time.  The \(l=1\) amplitude is amplified by self gravity.}
  \label{fig:early}
\end{figure}

After running the primary simulation for approximately 7 virial time
units (or 14 Gyr in Milky-Way time), the \(l=1\) power is distinctly
elevated.  The gravitational power is computed directly from the
biorthogonal expansion used to compute the N-body simulation as
described in Appendix \ref{sec:biorth}.  Figure \ref{fig:early} shows
the power for each harmonic order from equation (\ref{eq:power}) for
the primary simulation.  The self-gravity in the weakly damped mode
excited by the Poisson fluctuations elevates the power relative to the
\(l>1\) power.  This is described in \citet{Weinberg:94}.

\begin{figure}
  \includegraphics[width=0.45\textwidth]{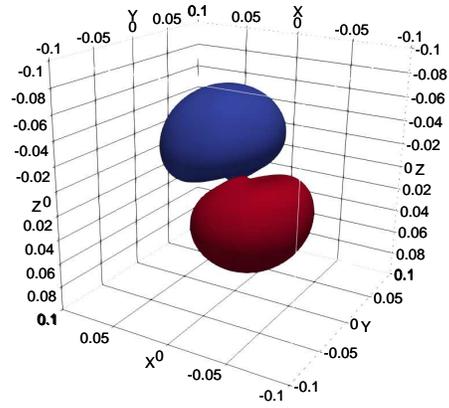}
  \caption{\label{fig:early_shape} The mSSA reconstruction of the
    \(l=1\) feature at \(T=5\) or roughly at 10 Gyr in Milky-Way
    units.  The isodensity surfaces illustrate the shape of the
    positive (red) and negative (blue) lobes of the dipole feature at
    the 10\% level relative to the peak. The overdensity relative to
    the background equilibrium is 0.25\% (\(T=5\)) at peak and this
    increases to 5\% at late times. \label{fig:mssa_render}}
\end{figure}

Then, we applied the mSSA technique to find the gravitationally
correlated signal for \(l=1\).  Approximately 95\% of the total
variance are in the first three eigenvalues.  A reconstruction of the
field is shown in Figure \ref{fig:early_shape}.  The scale of the
feature coincides with characteristic scale of the NFW-like halo,
\(r_a\) (eq. \ref{eq:NFW}), and its gravitational influence extends
beyond the halo half-mass radius.  Its shape is consistent with the
weakly damped modes investigated in
\citet{Weinberg:1994,Heggie.etal:2020,Fouvry.etal:2021}.

\begin{figure}
  \includegraphics[width=0.45\textwidth]{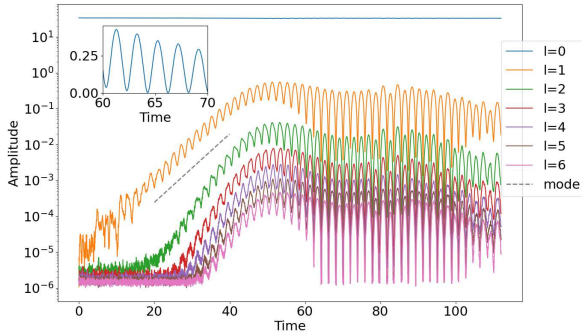}
  \caption{The amplitude for harmonic orders \(l=0,\ldots,4\) for the
    same run as in Fig. \ref{fig:early} extended to \(T=110\).  The
    inset shows the periodicity in the \(l=1\) power with a period of
    approximately 2 time units (or 4 Gyr in Milky-Way scaling).  The
    amplitude of the \(l=1\) mode relative to the background \(l=0\)
    is approximately 10\%. At late times, the low \(l\) power aliases
    into higher order spherical harmonics will similar shape and trend
    to \(l=4\) but successively lower amplitude.  We have omitted
    \(l>4\) curves for clarity.  The exponential growth rate predicted
    by linear response theory (dashed line) reproduces that in the
    simulation. \label{fig:full} }
\end{figure}

Initially, the mode does not have a constant pattern speed but appears
to be episodically fixed in space for several time units and then
moves quickly to a new orientation.  To investigate this strange and
unexpected behaviour, we ran the simulation much further to
\(t\lesssim100\) or approximately 30 Hubble times for Milky-Way
scaling.  The power plot for this extended run appears in Figure
\ref{fig:full}.  Several key features are readily apparent:
\begin{enumerate}
\item mSSA reveals that the dominant mode at \(T\lesssim5\) is a
  damped mode with \(\Re(\omega)\approx3.2\).  We confirm this with
  linear response theory in Section \ref{sec:linear}.  This suggests
  the initial amplitude of this mode is the projection of the \(l=1\)
  Poisson noise in the initial conditions.  The amplitude of this mode
  slowly decays in time (see Sec. \ref{sec:evidence}).

\item The unstable \(l=1\) mode is present and exponentially growing
  from the beginning of the simulation.  The real frequency of the
  growing mode is \(\approx1.9\).  The early-time behaviour is
  obscured by blending with the noise-driven damped mode, but the
  growing mode has a clear, coherent pattern beginning at
  \(T\approx10\).
\item The \(l=1\) power grows exponentially until \(T\lesssim40\).
\item Beyond this point, the \(l=1\) power is approximately constant
  and perhaps slowly decaying for \(T>40\).
\item The response shows a periodic modulation of approximately 2 time
  units (or 4 Gyr scaled to the Milky Way halo).
\item At early times, only the \(l=1\) power increases but later,
  \(T\gtrsim20\), all harmonics begin to grow.  This \(l>1\) power
  results from the non-linear saturation of the initially dipole-only
  mode aliased into higher-order harmonics.  There is no evidence from
  the mSSA analysis for new instabilities with \(l>1\).
\end{enumerate}
While surprising, this is not in violation of the Antonov stability
criterion.  The infinite-extent two-power NFW profile has
\(df(E)/dE<0\) but the truncated density distribution in equation
(\ref{eq:NFW}) does not.  The left-hand panel in Figure
\ref{fig:fEtrunc} shows the density profile from equation
(\ref{eq:NFW}) with width \(\sigma_t=n/10\) and \(r_t=1+3\sigma_t\)
for \(n=1,\ldots,7\).  The right-hand panels in this figure show the
isotropic DF from the Eddington inversion.  The lower right-hand side
panel is a zoom in of the \emph{bump} in the DF induced by the
truncation \({\cal T}\).  The truncation radius, \(r_t\), give rise to
a new energy scale, \(E_t\).  This results in an inflection in the
distribution function \(f(E)\) for \(E\sim E_t\).  The change in the
phase-space gradient near \(E_t\) enables a feedback channel between
the inner and outer halo that drives the instability
(Sec. \ref{sec:analysis}).  Even though we create this new scale by
radial truncation, a wide variety of natural processes in DM haloes,
such as accretion and disruption of dark subhaloes and dwarf galaxies
or incomplete mixing from the halo's early assembly history, can
introduce special scales that break the monotonic run in \(f(E)\) and
result in similar dynamics.  In summary, we use truncation as a device
to introduce a particular scale but expect the implications to be
generic.  For example, we demonstrate in Section \ref{sec:linear} that
anisotropy causes an inflection in the distribution function which
also results in instability.

We speculate that the existence of weakly damped and weakly growing
\(l=1\) rather than \(l>1\) modes has several related origins.  First,
the damping rate in the point modes tends to increase with harmonic
order because higher harmonic order presents more opportunities for
commensurabilities at higher frequencies.  Secondly, the
self-gravitating pattern formed from a group of slowly precessing
eccentric orbits is anomalously low compared to the azimuthal orbital
frequencies.  It therefore couples weakly to surrounding phase space.
Finally, the low \(l=1\) modal frequency circumstantially coincides
with the characteristic orbital frequency in the outer halo coinciding
with the \emph{bump} in the distribution function.  This provides an
opportunity for instability.

\section{Analysis and explanation}
\label{sec:analysis}

\subsection{Evidence for an \(l=1\) instability from N-body
  simulations}
\label{sec:evidence}

The phase-space distribution for spherical systems with no preferred
axis may be represented by two action variables, \((I_1, I_2)\). The
radial action is \(I_1\) and the orbital angular momentum is
\(I_2\). The third action, \(I_3\), is the azimuthal component of the
orbital angular momentum, \(J_z\).  Let the conjugate angles be
\((w_1, w_2, w_3)\).  Alternatively, we may describe phase space by
independent functions of the actions.  Here, we use energy
\(E=E(I_1, I_2), \kappa = I_2/J_{\mbox{\tiny max}}(E)\) where
\(J_{\mbox{\tiny max}}(E)\) is the angular momentum for the circular
orbit with energy \(E\).  The variables \((E, \kappa)\) provide a
convenient rectangular domain.

\begin{figure}
  \includegraphics[width=0.5\textwidth]{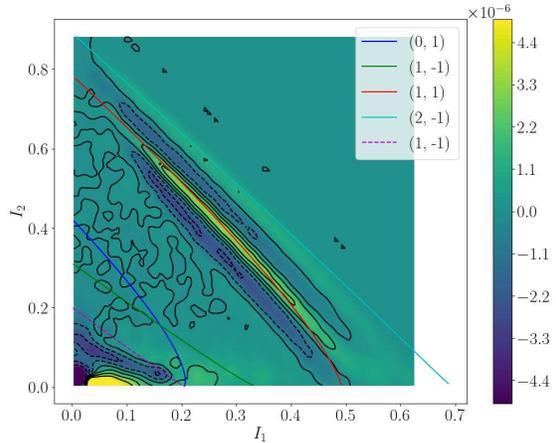}
  \caption{The change in the phase-space distribution function,
    \(\Delta f(\mathbf{I})\), in radial actions and orbital angular
    momentum (\(I_1, I_2\)) during the initial exponential growth
    phase.  The zero-amplitude location of the primary
    commensurabilities with
    \(l_1\Omega_r + l_2\Omega_\phi = \Omega_p\) are shown as colour
    coded curves labeled by \((l_1, l_2)\).  Solid (dashed) lines
    correspond to \(\Omega_p=1.82\) (\(3.21\)).  These are the
    predicted pattern speeds of the growing (damped) modes. The large
    changes in \(f(\mathbf{I})\) at small values of \((I_1, I_2)\) in
    the lower left-hand corner are distortions in the equilibrium
    model driven by the growing mode.  The resonant couples advect
    orbits to larger \(I_1\) and smaller \(I_2\), primarily through
    the \((1, -1)\) resonance. 
    \label{fig:I10}}
\end{figure}

Figure \ref{fig:I10} shows the change the distribution function in
action space between between the time \(T=0\) (initial) and time
\(T=10\), the beginning of exponential growth phase.  To estimate the
distribution function, we bin the actions on a \(60\times60\) grid in
\((I_1, I_2)\) and normalise to unit mass and phase-space volume.  We
then compute the difference between the initial and final densities
averaged over 8 separate simulation snapshots.  This extra bit of
averaging reduces the sampling noise.  All harmonic orders of the
response are represented, not only \(l=1\), but the \(l=1\)
interaction dominates.  The location of the primary low-order
commensurabilities for \(\Omega_p\approx1.82\) and
\(\Omega_p\approx3.21\) also are shown in Figure \ref{fig:I10}. These
two point modes are identified by linear perturbation theory using the
methods described in Section \ref{sec:linear} and Appendix
\ref{sec:disper}).  We will demonstrate below that these two
frequencies correspond to the exponentially growing and first damped
mode, respectively.  We denote the commensurabilities by the pair of
integers \((l_1, l_2)\) such that
\(l_1\Omega_1 + l_2\Omega_2 = \Omega_p\) where \(\Omega_1, \Omega_2\)
are the radial and tangential frequencies, respectively. The resonance
tracks in \(I_1, I_2\) appear as anti-diagonal loci, generally.

Figure \ref{fig:I10} shows that the global response is dominated by
the \((1, 1)\) resonance.  This is resonance overlaps the position of
the inflection in the background distribution function and transports
energy and angular momentum from the point modes to larger values.
The prominent feature between \(I_1\in[0,0.2], I_2\in[0.05,0.1]\)
results from a damped mode rather than the growing mode.  The damped
mode couples through the \((1, -1)\) at early times shown in this
figure. This mode also results in transport from lower to higher
energy and angular momentum.  At later times, this couple disappears
as the damped mode fades.  This mode may be excited by the slight
disequilibrium of the initial conditions, by collective relaxation
driven by particle noise, or perhaps both.  Further precise
identification of its origin is elusive.  The disappearance of the
damped mode at late times suggests an excitation by the initial
disequilibrium.  However, the growing mode changes the underlying
structure of the distribution, and this could modify the dressed,
collective response.

\begin{figure}
  \includegraphics[width=0.45\textwidth]{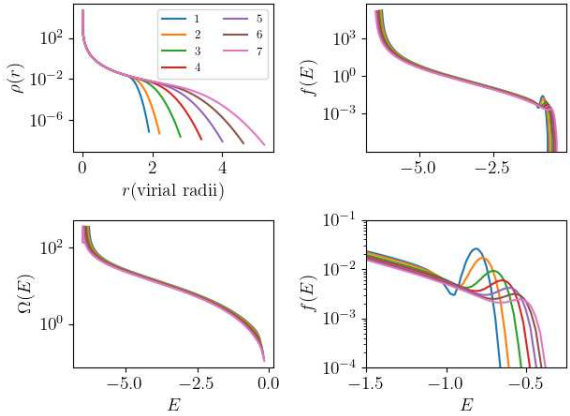}
  \caption{Upper left: the truncated density profile from
    eq. \ref{eq:NFW} for the truncation index \(n\).  Smaller index
    denotes sharper truncation.  Lower left: the circular orbit
    frequency as function of energy for the same models. a Right: the
    isotropic distribution function, \(f(E)\), for various truncation
    widths. The sharper the cut off, the larger the inflection in the
    run of \(f(E)\).  This bump in the tail of the distribution
    function powers the \(l=1\) instability.  The upper panel shows
    the full range of \(f(E)\) and the lower panels zooms in on the
    bump.  \label{fig:fEtrunc}}
\end{figure}

Rather than employing linear response theory to investigate this
intrinsically non-linear development, we use mSSA to explore the
evolution empirically.  The mSSA analysis applied to the \(l=1\) time
series of coefficients for our fiducial simulation provides a series
of principal components in order of contribution to the temporal
variance.  While mSSA can approximate a signal with changing
frequencies, it will also tend to separate signals of different
frequencies into multiple PC pairs depending on the duration of the
interval at a particular frequency.  For example, imagine a time
series that begins with a constant oscillation at frequency \(\nu_0\),
changes its frequency over a short period in the middle of series to a
new frequency \(\nu_1\) with \(\nu_1\ll\nu_0\), followed by a constant
oscillation at \(\nu_1\) at late times.  MSSA will provide two
dominant PC pairs describing the early and late times with a third
pair describing the transition.  Conversely, mSSA will nicely separate
a sinusoidally modulated chirp into to distinct PCs, one for the
modulation and one for the chirp, as described in
\citet{Weinberg.Petersen:2021}.  In summary, mSSA will separate
punctuated signals into discrete intervals.  Similarly, a DFT will
characterise the frequency features in a PC but provide no information
on where what time interval in the series dominates.  So the two
methods together are complementary.  See Section \ref{sec:saturated}
for an in-depth characterisation of the saturated mode using this
approach.

\begin{figure}
  \includegraphics[width=0.5\textwidth]{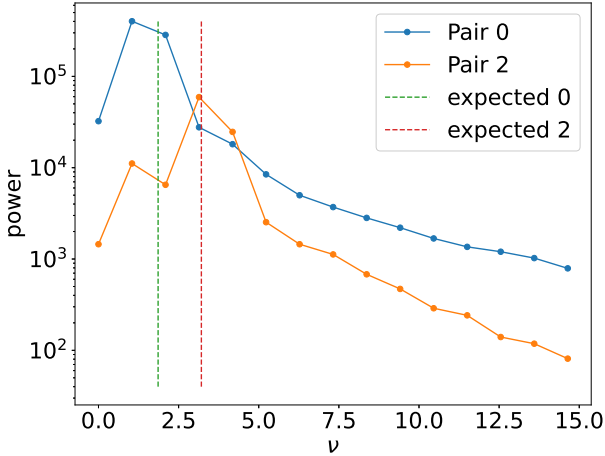}
  \caption{The DFT power from two dominant PC pairs from the mSSA
    analysis of \(l=1\) coefficients with \(n_{\mbox{\tiny max}}=12\).
    Only the low frequency part of the spectrum is shown; the power at
    larger \(\nu\) decrease exponentially with increasing \(\nu\).
    The dashed vertical lines show the real frequencies of the
    exponentially growing mode and first damped mode from linear
    response theory for comparison. \label{fig:pc_dft}}
\end{figure}

For an example pertinent to the results of this and the next section,
we apply mSSA to the \(l=1\) coefficients in the same time interval,
\(T\in[0,12]\), used for Figure \ref{fig:fEtrunc}.  Figure
\ref{fig:pc_dft} describes the power from two low-order PCs pairs from
\(l=1\).  The first pair (labeled `Pair 0') describes most of the
\(l=1\) power in Figure \ref{fig:full}.  The third pair (labeled `Pair
2') corresponds to the damped mode that dominates at early times.  The
frequency of peaks in each pair matches the the pattern speeds
predicted by linear theory (see Sec. \ref{sec:linear}).  Pair 1, not
shown, is a blend of both frequencies.  In summary, our application of
mSSA to the N-body simulation recovers the main features of the
expected linear development.  However, we will see in the following
sections that mSSA provides useful characterisation in the non-linear
regime as well.

\begin{figure}
  \includegraphics[width=0.5\textwidth]{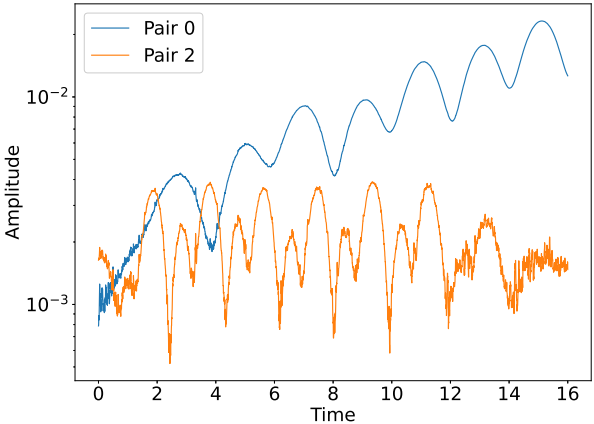}
  \caption{The coefficient amplitude (as in Fig. \ref{fig:early}) from
    two dominant PC pairs from the mSSA analysis of \(l=1\)
    coefficients with \(n_{\mbox{\tiny max}}=12\). \label{fig:pc_amp}}
\end{figure}

The power in the mSSA-derived PCs is power in the detrended,
normalised coefficient time series.  These DFT power spectra are
\emph{not} physical power in the gravitational field but the
correlated temporal variation in the coefficient series about their
detrended values.  This \emph{detrending} is motivated by the
mathematics of SSA.  The method will cleanly separate individual
signals from a dynamical process that can be represented by a linear
recurrence relation \citep[see][Section 2.2 and Chapter
5]{Golyandina.etal:2001}.  For example, exponential and
trigonometric functions are exactly described by first- and
second-order recurrences.  The detrended BFE coefficient series look
like zero-mean, unit-variance signals, and provide better separation;
mSSA finds the growing, rotating mode clearly, isolating most of the
feature in a pair of eigenvectors.  For physical interpretation, we
reconstruct these PCs into BFE coefficients.  The amplitude of the
coefficients for the same two PC pairs in Figure \ref{fig:pc_dft} is
shown in Figure \ref{fig:pc_amp}.  This figure shows that the
gravitational energy in Pair 0 from Figure \ref{fig:pc_dft} is
exponentially growing and the power in Pair 2 is slowly decaying.

\subsection{The saturated mode}

\label{sec:saturated}

The exponential growth saturates at \(T\approx40\) (see
Fig. \ref{fig:full}) and the behaviour of the mode changes.  The
radial phase of participating eccentric orbits correlate, and the
orbits sustaining the mode tend to bunch near apocentre.  This is
similar to the behaviour of the radial orbit instability: the
precession angles bunch near apocentric angle yielding a quadrupole
distortion.  In this case, the orbits bunch in angle and oscillate in
phase between pericentre and apocentre causing the modulation of the
dipole power seen in Figure \ref{fig:full}.  The shape of the
non-linear mode in these panels is similar to the early-time
reconstruction in Figure \ref{fig:early_shape}.  While the mode is
sustained by the gravitational field of orbits strongly bunched in
radial phase, the overall response of the halo is similar to the
early-time feature which is only weakly bunched in phase.

\begin{figure}
  \includegraphics[width=0.50\textwidth]{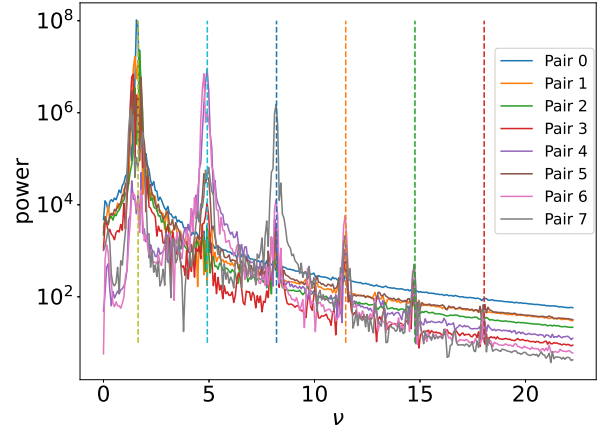}
  \caption{The DFT power of the first eight PC pairs for the mSSA
    analysis of the BFE coefficients beyond the saturation phase of
    the simulation (cf. Fig. \ref{fig:full}). Here the dashed vertical
    lines show the odd frequency harmonics of the fundamental
    frequency with \(\omega\approx1.64\).
    \label{fig:sat_dft}}
\end{figure}

\begin{figure}
  \includegraphics[width=0.45\textwidth]{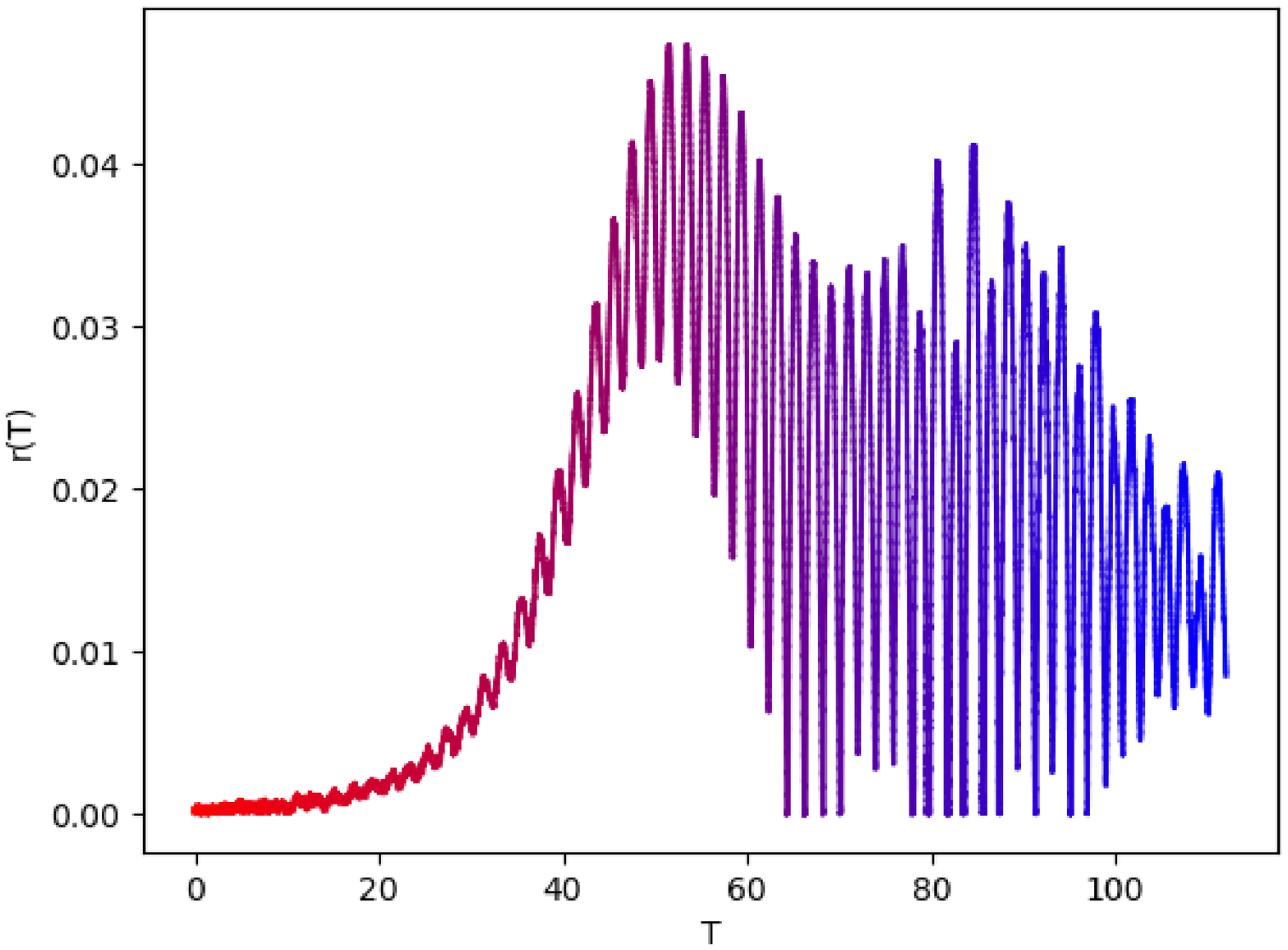}
  \includegraphics[width=0.50\textwidth]{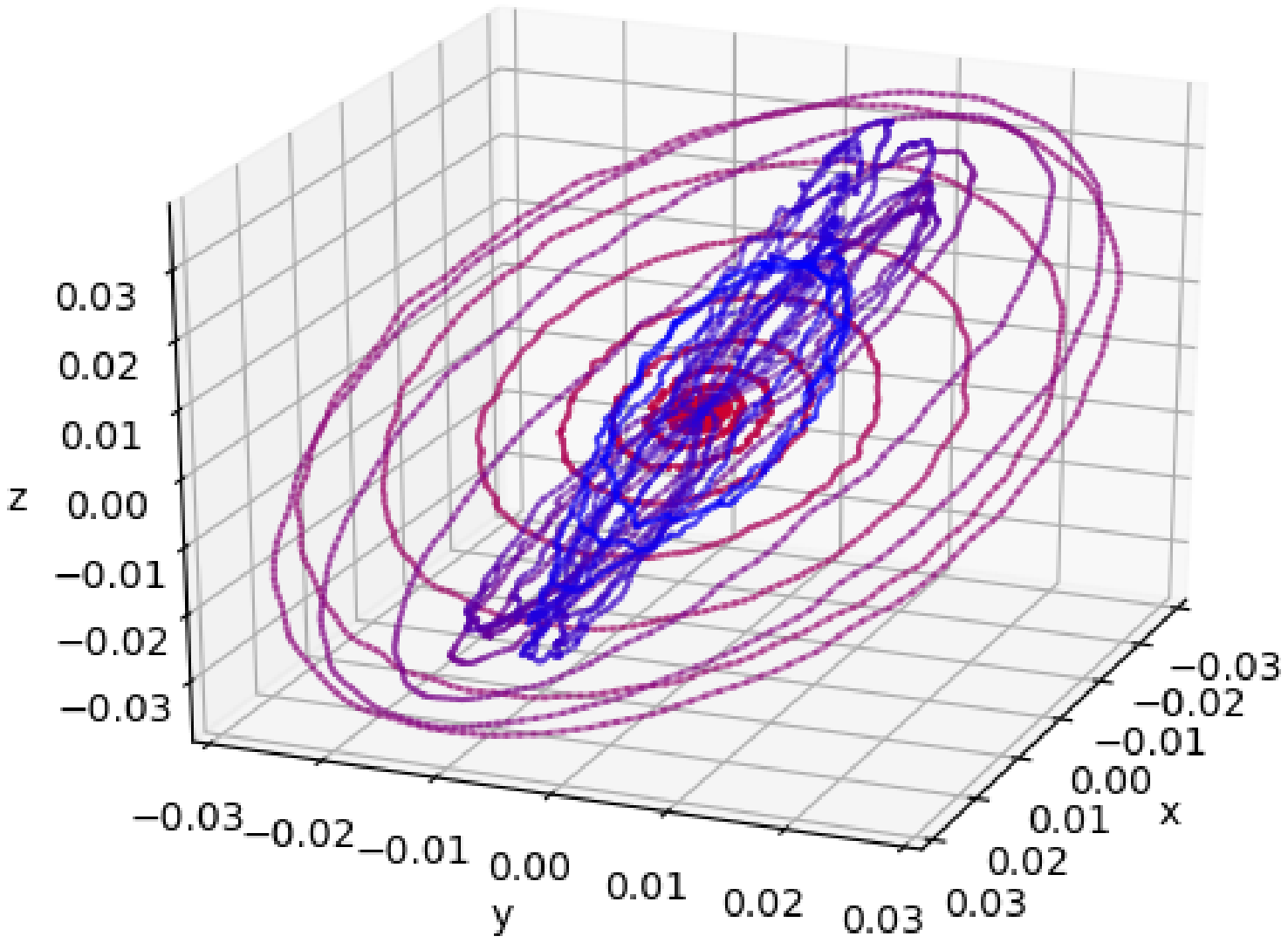}
  \caption{Top: the radial displacement of the cusp from the geometric
    centre of the halo as a function of time.  Bottom: location of the
    peak density centre in three-dimensional space; time is colour
    coded as in the top panel.  The orientation has been chosen
    roughly face-on to best illustrate trajectory of the cusp induced
    by the dipole mode.  At early times (\(T\lesssim20\), we see
    oscillation with a uniform pattern speed with an slow outward
    spiral, consistent with the expected the linear behaviour of the
    mode.  For \(20\lesssim T\lesssim40\), the non-linear bunching
    develops and the trajectory of the centre becomes more eccentric.
    For \(T\gtrsim40\), the trajectory becomes eccentric, the cusp
    begins to oscillate along a diameter.
    \label{fig:cusp}}
\end{figure}

The details of the transition from linear to non-linear are as
follows. As the mode grows, its period increases. For the interval
\(T\in[0,10]\), one PC has the predicted frequency of the linear mode,
\(\Re(\omega)\approx1.8\), and one has the faster damped-mode
frequency, \(\Re(\omega)\approx2.5\) as described in Figure
\ref{fig:pc_dft}.  At early times, e.g. the interval \(T\in[0,5]\),
the damped linear-mode frequency is dominant.  As the interval
increases from \(T\in[0,20]\) to \(T\in[0,40]\), the longer-period
exponentially growing oscillation dominates.  In the interval
\(T=[0, 20]\), the first PC pair describes the exponentially growing
mode with frequency \(\Re(\omega)\approx1.8\).

The DFT of the PCs for the larger interval, \(T=[0, 40]\) is shown in
Figure \ref{fig:sat_dft}.  The peak in the same dominant PC pair
becomes narrower and the initial damped-mode frequency is absent; the
disturbance is nearly completely dominated by the non-linear mode for
\(T>20\).  As the non-linear bunching in phase angle develops, the
pattern speed slows to \(\Re(\omega)\approx1.6\).  This suggests that
the response mode begins close to the linear prediction: a constant
pattern speed with exponential growth.  As the amplitude increases and
the disturbance becomes non-linear, the pattern becomes time dependent
and shifts to a lower frequency typical of its nearly resonant orbits.
If the mode were purely sinusoidal, the total power for the PC pairs
would show \emph{only} the slow change in overall amplitude.  The
strong modulation of the amplitude at late times in Figure
\ref{fig:full} indicates that the mode's oscillation behaviour is
similar to that of a single nearly radial orbit consistent with
bunching.  Quantitatively, this results in strong odd harmonics of the
primary frequency in the spectrum (Fig. \ref{fig:sat_dft}).  For
comparison, the figure marks the location of the odd harmonics of
fundamental frequency: \(\omega\approx1.64\).

At the same time, the gravitational potential from the dipole
oscillation begins to affect the location of the inner cusp although
the inner profile of the cusp itself remains intact.  The internal
dynamics of the cusp itself is decoupled from that of the mode but does
provide a convenient diagnostic for the global effect of the dipole.
The location of the cusp is depicted in Figure \ref{fig:cusp}.  At
early times, the dipole rotates with a uniform pattern speed as
expected from linear theory (see Sec. \ref{sec:linear}).  Combined
with the growth in amplitude, the trajectory of the cusp is an
outward-moving spiral that follows the pattern of the linear mode.  At
later times, \(T\gtrsim40\), the cusp trajectory becomes increasingly
radial along a diameter.  This characterisation provides evidence that
the saturated mode is an ensemble of radially-biased orbits with
similar apocentre (and therefore energy) and is consistent with the
rapid radial phase reversal typical of a nearly radial orbit.  When
this transition occurs, the resonance between the modal pattern and
the outer halo decouples. In addition, Figure \ref{fig:cusp} shows
that the principal plane of the mode remains fixed in space, as it
must to conserve net angular momentum.  The orientation of the
diameter in this simulation is random.

\begin{figure}
  \includegraphics[width=0.5\textwidth]{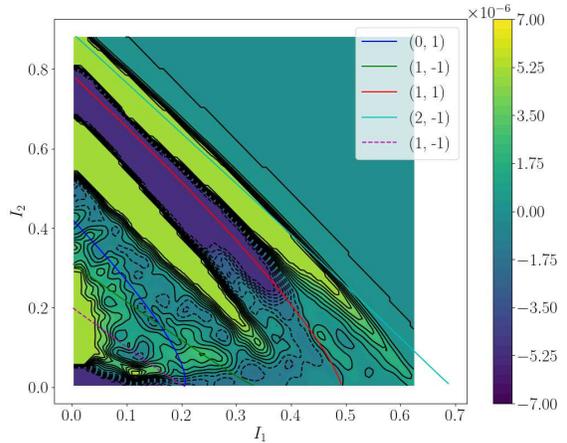}
  \caption{The change in the phase-space distribution function as in
    Fig. \ref{fig:I10} but shown near the peak amplitude in the
    saturated phase, \(T\approx48\).  The loci of the unperturbed
    resonances from Fig. \ref{fig:I10} are shown for comparison.
    \label{fig:I11}}
\end{figure}

Figure \ref{fig:I11} describes the change in the distribution function
from exponential growth at \(T=36\) to saturation at \(T=48\) (cf.
Fig. \ref{fig:I10}).  The change in the phase-space distribution
function is characterised by a flat-bottomed valley surrounded by two
flat-topped ridges in roughly the same location as the
\((l_1, l_2) = (1, -1)\) feature in Figure \ref{fig:I10}.  The change
in \(I_1, I_2\) in the lower left-hand corner of Figure \ref{fig:I11}
corresponding to \(I_1\lesssim0.05, I_2\lesssim0.3\) is caused by the
motion of the inner profile as described in Figure \ref{fig:cusp}.
These ridge and trough features in the DF suggest that the resonant
couple has also saturated.  Future work will be necessary to
demonstrate whether the origin of the saturation is the radial-angle
bunching which detunes the resonance, the secular modification of the
bump in the phase-space distribution function, or the interaction of
both.

\subsection{Combined disc and halo models}

This \(l=1\) mode will be modified by the gravitational field of the
combined disc and halo.  Although difficult with the Hamiltonian
perturbation theory used in Section \ref{sec:linear}, a numerical
investigation is straightforward with the BFE+mSSA approach.
Preliminary results for the evolution of a disc and halo indicates
that the \(l=1\) halo feature has the same scale and magnitude as
described earlier.  However, the oblate potential induced by the disc
component causes the \(l=1\) pattern to nod in the disc plane on an
approximately Gyr time scale.  This will affect the luminous stellar
disc inevitably.  This mode could be a natural source of lop-sided
asymmetries in Nature and will be the subject of future research.

\section{Verification of the instability}
\label{sec:verify}

We anticipate that some readers will be concerned that the
exponentially growth is an artefact of using a BFE method rather than
a direct- or tree-gravity code. For example, the BFE technique
requires an expansion centre, and this centre may bias the force in
such a way to produce a false instability, perhaps?  We tested this in
two ways:
\begin{enumerate}
\item \EXP\ can centre the expansion on the minimum of the
  gravitational potential by ranking the particles by gravitational
  potential and computing the centre of mass of the 1000 most bound
  particles.  This option is designed for live satellite simulations;
  the default simulation does not use this option.  To test the effect
  of centring, we resimulated the fiducial run with and without this
  centring option.
\item We used \GADGET\ on the same particle distribution with softening
  length designed to minimise the force error as described in
  \citet{Dehnen:2001}.  We used a kd-tree neighbourhood search and
  kernel density estimation to estimate the local density at each
  particle position and chose the expansion centre to be the centre of
  mass for the 1000 highest density particles positions.  We then used
  \EXP\ to compute the expansion coefficients from the \GADGET\
  snapshot files.  The same centring scheme was used for both \EXP\
  and \GADGET\ phase-space output for direct comparison.
\end{enumerate}

The results of Test (i) are simple: the results were nearly the same.
This further implies that the number of radial basis functions are
sufficient to reproduce the density centre displacement in these \EXP\
simulation.  See Appendix \ref{sec:center} for additional discussion
of centring.

\begin{figure}
  \includegraphics[width=0.45\textwidth]{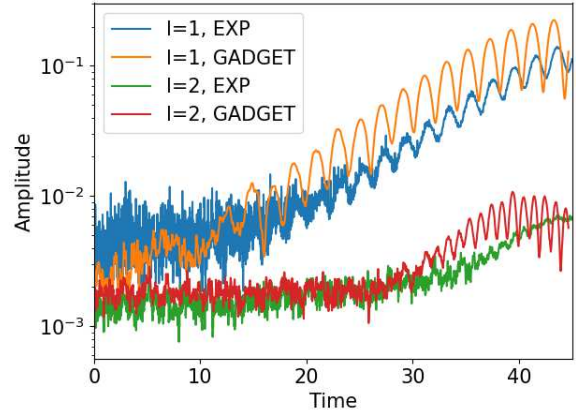}
  \caption{Gravitational amplitude for harmonic orders \(l=1,2\) from
    the \GADGET\ and \EXP\ simulations and using the same particle
    distribution from Fig. \ref{fig:full}.  The expansion centre for
    each is computed by density estimation from kd-tree neighbourhood
    search.}
  \label{fig:gadget}
\end{figure}

The results of Test (ii) require additional computation to make a fair
comparison.  The resulting power estimates will be different from the
ones in Figure \ref{fig:full} because the density centre is not fixed
in inertial space but moving under the influence of the mode and
noise.  In addition, the centres for the two simulations are now
different owing to systematics in the numerical Poisson solvers,
although one expects the peak density and minimum potential to be
correlated.  Nonetheless, the comparison between the \(l=1, 2\) power
computed from the \EXP\ and \GADGET\ simulations using the kd-tree
centring scheme for each (Fig. \ref{fig:gadget}) demonstrates
reasonable qualitative and quantitative agreement.  Specifically, the
\(l=1\) shows the same exponential growth rate and the modulation has
the same period as the fiducial \EXP\ run.  Also, the aliased \(l=2\)
power approximately coincides suggesting that this power is a centring
artefact itself.  A detailed examination of the centre location
reveals that the mean offsets are approximately 0.002 virial units (or
600 pc in Milky Way units) which is very small compared to the average
scale of the mode which is 0.05 (or 15 kpc in Milky Way units).  It
seems safe to conclude that the unstable mode is not an artefact of
our BFE-based Poisson solver.

\section{Insights from linear-response theory}
\label{sec:linear}

\begin{figure}
  \includegraphics[width=0.45\textwidth]{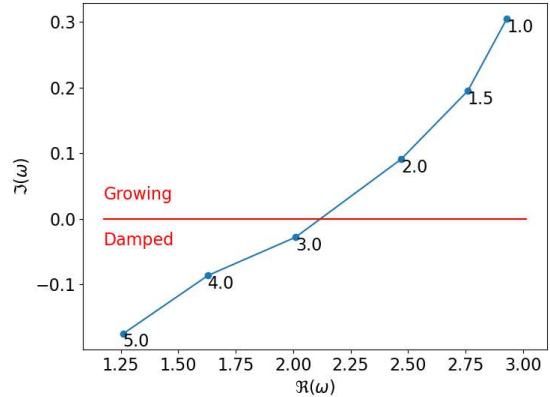}
  \caption{The location of zeros the \(\Re(\omega)\)--\(\Im(\omega)\)
    plane for the fiducial model with an isotropic distribution
    function \(f(E)\) as a function of truncation index \(n\) which
    defines the width of the truncation function \(\sigma_t = n/10\)
    (see Sec. \ref{sec:instability}).
    \label{fig:mw_l1_trunc}}
\end{figure}

This section applies linear response theory used in
\citet{Weinberg:94} and described in
\citet{Kalnajs:77a,Fridman.Polyachenko:84,Binney.Tremaine:2008} to
corroborate the N-body and numerical analyses of the previous
sections.  In short, we seek the simultaneous solutions of the
collisionless Boltzmann and Poisson equations.  By representing the
density and potential fields by the BFE previously described, this
method discretises function space, reducing the solution for modes to
the zeros of a determinant, called the \emph{dispersion relation}.
Because this discretisation results in finite-dimension approximation
by a matrix, it is often called the \emph{matrix method}.  The linear
response code uses the same algorithms as in
\citet{Weinberg:89,Weinberg:94} although the code has been updated to
modern {\tt C++-17}.  Solutions with \(\Im(\omega)>0\)
(\(\Im(\omega)<0\)) are growing (decaying).  The derivation and
evaluation is described in Appendix \ref{sec:disper}.  As a basic
check, we verified that there are no growing modes for \(f=f(E)\) in
the untruncated NFW model as demanded by the Antonov theorems.  Also,
we checked the results quoted for the \(l=2\) instabilities in
\citet{Weinberg:89} were recovered.

The goal of this section is not to further elucidate the simulations
previously described but to demonstrate that the \emph{bump on tail}
instability identified in Section \ref{sec:instability} is not
dependent on the specifics of our fiducial model
(Sec. \ref{sec:models}).  We explore two strategies for modifying the
DF to cause a \emph{bump}.  Section \ref{sec:isotrunc} describes the
response frequencies for isotropic distribution function with varying
degrees of truncation width, \(\sigma_t\) (eq. \ref{eq:trunc}).
Section \ref{sec:qra} explores varying anisotropy radii, \(r_q\), for
truncation at 2.5 virial radii.  In both cases, the result of
truncation and increasing anisotropy is an inflection in the run of
\(df(E)/dE\) for the isotropic DF and \(df(Q)/dQ\) for the
Osipkov-Merritt DF.  The resonant couples that drive instability
depend on the phase-space and model structure in detail and we do not
have an explicit criteria for the \(l=1\) stability other than the
numerical evaluation of the dispersion relation itself.  Nonetheless,
the examples here suggest that the bump in the DF is the cause.

\subsection{Isotropic NFW-like models with varying truncation}
\label{sec:isotrunc}

We apply the linear response theory to the isotropic distribution
functions resulting from Eddington inversion of equation
(\ref{eq:NFW}) for varying truncation width \(\sigma_t\).  We choose
widths \(\sigma_t=n/10\) for \(n=1,\ldots,7\) and the truncation radii
\(r_t=1+3\sigma_t\) as described in Section \ref{sec:instability} to
ensure that the model remains close to the NFW profile at the virial
radius and rolls over only at larger radii. The location of the
dominant response mode frequencies for \(l=1\) is illustrated in
Figure \ref{fig:mw_l1_trunc}.  The models are \(l=1\) unstable for
small values of \(\sigma_t\).  By increasing the width of the
truncation, the mode crosses the real axis into stability.  See Figure
\ref{fig:fEtrunc} for a graphical depiction of these profiles.

The fiducial simulation described in Sections \ref{sec:instability}
and \ref{sec:analysis} has \(n=2\).  For reference, the index \(n=2\)
corresponds to a truncation width of 1/5 virial radius and truncation
radius at 1.6 virial radii.  The index \(n=5\) corresponds to a
truncation width of 1/2 virial radius and truncation radius at 2.5
virial radii.  Figure \ref{fig:mw_l1_trunc} corroborates results of
previous sections: the fiducial model is \(l=1\)
unstable\footnote{This series of models has a slightly different
  linear scale and therefore a slightly different pattern speed than
  the fiducial model from Section \ref{sec:instability}}.

This instability appears to result from a bump in the tail of the
distribution with \(df(E)/dE>0\).  The evidence is as follows. The
pattern speed, \(\Re(\omega)\), of the weakly growing mode is
\(\sim1.9\) for the values of \(n\) considered (see
Fig. \ref{fig:mw_l1_trunc}). As described in Section
\ref{sec:instability}, this frequency decreases to \(\sim1.6\) as the
mode grows in amplitude. The pattern frequency matches the orbital
frequencies at the same energy as the bump in the DF induced by the
truncation (see Fig. \ref{fig:fEtrunc}), suggesting the involvement of
a low-order resonance.

\begin{figure}
  \includegraphics[width=0.45\textwidth]{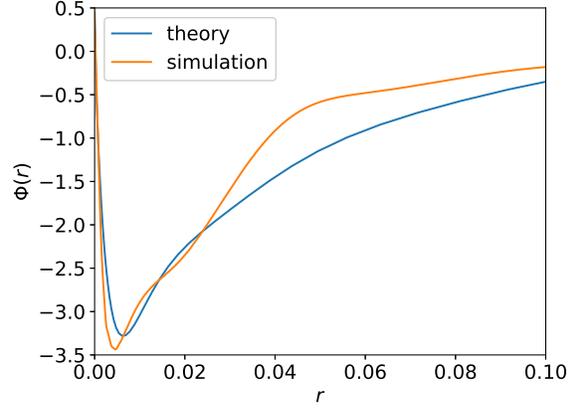}
  \caption{\label{fig:mode_compare} The first PC pair from the mSSA
    analysis from Sec. \ref{sec:instability} at \(T=10\) provides the
    radial profile of the \(l=1\) dipole distortion (denoted as
    \emph{simulation}).  This is compared with the shape of the unstable mode
    inferred from the linear response analysis (denoted as
    \emph{theory}).  The overall scaling in amplitude is arbitrary but
    the shape and position fully determined by the simulation and
    linear theory.}
\end{figure}

The linear analysis predicts the eigenfunction (see Appendix
\ref{sec:modal_response}) once the frequency is identified.  The shape
and extent are consistent with the profile inferred from the
simulation (Fig \ref{fig:early_shape}) and the two are quantitatively
compared in Figure \ref{fig:mode_compare}.  This provides further
corroboration that the identification of the mode from linear theory
is correct.  Although we do not expect the linear theory to represent
the profile at late times, after non-linear saturation, the
qualitative features remain similar.

\subsection{Isotropic NFW-like models with varying outer slope}
\label{sec:beta}

\begin{figure}
  \includegraphics[width=0.45\textwidth]{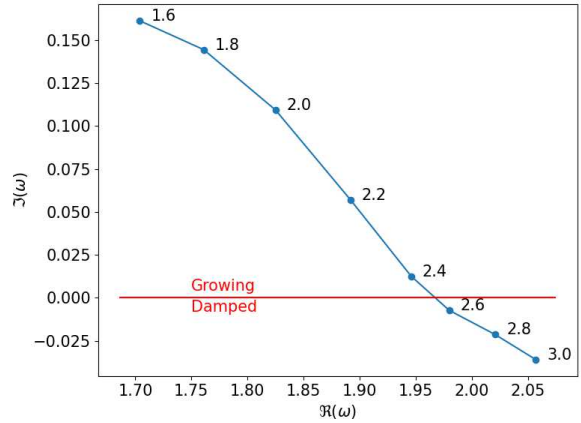}
  \caption{The location of zeros on the
    \(\Re(\omega)\)--\(\Im(\omega)\) plane for the fiducial model as a
    function of the outer power-law index \(\beta\) (labeled) with the
    same truncation as the fiducial model.  The density
    \(\rho(r)\propto r^{-1-\beta}\) at large \(r\). The \(\beta=2\)
    model is the NFW profile and the \(\beta=3\) model is the
    Hernquist profile. \label{fig:mw_l1_beta}}
\end{figure}

We set the truncation parameter defined in Section \ref{sec:isotrunc}
to \(n=2\) and vary the outer slope \(\beta\) in equation \ref{eq:NFW}
with \(\beta\in[1.6, 2.8]\).  We fix the inner power-law index to
\(\alpha=1\), and therefore the density profile for \(r>r_a\) has
\(\rho(r)\propto r^{-1-\beta}\).  The zeros of the dispersion relation
for this sequence is shown in Figure \ref{fig:mw_l1_beta}.  As the
density profile steepens, the \(l=1\) mode becomes damped.  The
transition between growing and damped for this truncation occurs at
\(\beta\approx2.5\), between the NFW and Hernquist profiles.  While
researchers often prefer to work with the \(\beta=3\) Hernquist model
because it has finite mass, the frequency distribution of orbits will
be different between \(\beta=2\) and \(\beta=3\) and this will affect
the dynamical response. In this example, the NFW-like model is
unstable and the Hernquist-like model is stable.  This difference is
likely to also affect other astronomically relevant responses, such as
satellite wakes.  We urge researchers to consider their choice of halo
profile carefully and investigate the effect of the outer halo
profile.

\subsection{Anisotropic NFW-like models}
\label{sec:qra}

\begin{figure}
  \includegraphics[width=0.45\textwidth]{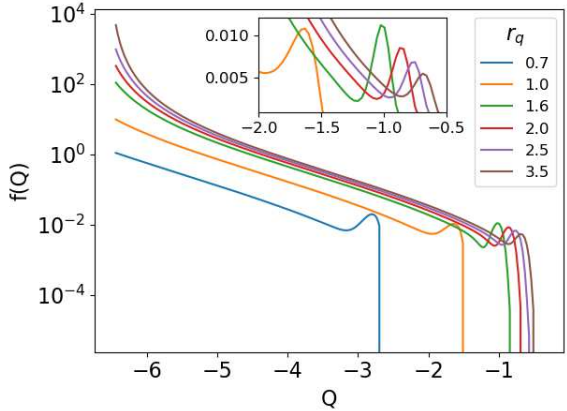}
  \caption{\label{fig:distfq} Anisotropic distribution function shown
    as a function of generalised energy, \(Q=E+L^2/2r_q^2\), for
    various values of anisotropy radius \(r_q\).  We show a value of
    \(r_q\) approaching the limiting value for reference; this extreme
    model is not investigated in this study.}
\end{figure}

\begin{figure}
  \includegraphics[width=0.45\textwidth]{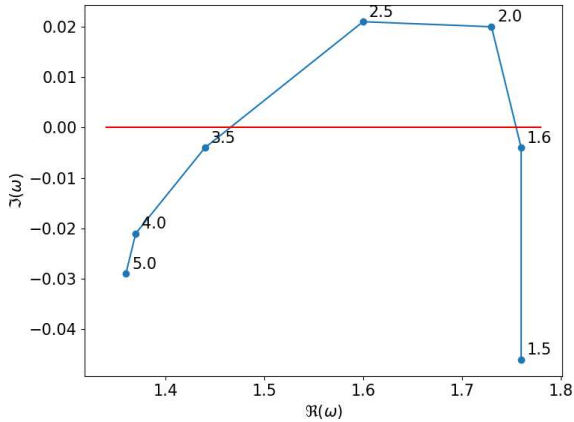}
  \caption{The location of zeros on the \(\Re(\omega)\)--\(\Im(\omega)\)
    plane for the fiducial model as a function of Osipkov-Merritt
    anisotropy radius \(r_q\). Models with a small range of values,
    \(1.6\lesssim r_q\lesssim 3.4\), are unstable.
    In all of these models, the outer halo is
    modestly anisotropic beyond virial radius and remains isotropic
    at smaller radii.  More importantly, the orbital frequency at
    location of the induced inflection corresponds to 
    the frequency of the primary mode.
    \label{fig:mw_l1_rq}}
\end{figure}

We examine the influence of anisotropy using the Osipkov-Merritt
generalisation of the Eddington inversion method for the same density
model as in the previous section with \(n=5\).  While these
distribution functions are not good descriptions of DM distributions
in cosmological simulations, this distribution function provides a
single parameter: the anisotropy radius, \(r_q\).  Section
\ref{sec:isotrunc} shows that the model is stable in isotropic limit,
\(r_q=\infty\).  Although one can bias the anisotropy towards both
radial and tangential, the radially anisotropic branch yields larger
changes in the DF and we restrict our attention to radial anisotropy.
The Osipkov-Merritt distribution function has a minimum value of
\(r_q\).  For smaller values than this critical value, the
distribution function becomes unphysically negative at small energies
(i.e. most bound orbits).  The critical value for this truncated
NFW-like model is \(r_q\approx0.5\).  The run of DF for a variety of
\(r_q\) values are illustrated in Figure \ref{fig:distfq}.  As \(r_q\)
decreases, the high \(Q=E+L^2/2r_q^2\) phase space is truncated.  This
implies that the higher angular momentum orbits are suppressed at high
energy; in other words, particles at large radii are eccentric and
have much smaller guiding centre radii.  This suppresses the lowest
frequency orbits.

The zeros of the dispersion relation in the complex \(\omega\) plane
for \(l=1\) are shown in Figure \ref{fig:mw_l1_rq}.  We find that
\(l=1\) modes are unstable for \(1.6\lesssim r_q\lesssim 3.4\).  This
is precisely the energy regime where the bump in \(f(Q)\) from Figure
\ref{fig:distfq} overlaps with orbital frequencies at similar values
to the pattern speed of the mode, \(\Re(\omega)\).  For smaller values
of \(r_q\), the frequencies of orbits corresponding to energies in the
bump are higher than the natural frequency of the mode, and the mode
damps.  As \(r_q\) increases, \(\Re(\omega)\) increases and
\(\Im(\omega)\) decreases; the value approaches that in Figure
\ref{fig:mw_l1_trunc}.  This anisotropy radius needed for growth is
large in model units, two virial radii, suggesting that a very minor
amount anisotropy producing a modest bump in \(f(Q)\) is sufficient
for \(l=1\) instability.

As the radial anisotropy approaches approximately 1.5, a different
unstable \(l=1\) modal track appears along with the \(l=2\)
ROI. However, the main goal of this investigation is not the dynamical
role of anisotropy per se, but a demonstration that the bump in the DF
drives the instability.  So we will not explore this alternate \(l=1\)
track here.

\begin{figure}
  \includegraphics[width=0.45\textwidth]{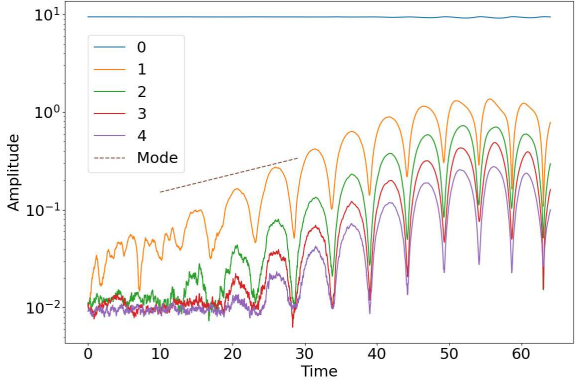}
  \caption{Growth in power for the NFW model with an Osipkov-Merritt
    radius \(r_q=2.0\), as described in Figure \ref{fig:full} for the
    fiducial model.  The growth rate predicted by the linear theory
    (dashed line labeled \emph{Mode}) is approximately 30\% smaller
    than in the simulation. Qualitatively, the evolution unchanged
    from the truncated, isotropic fiducial
    model. \label{fig:power_2p0} }
\end{figure}

We then simulated this same phase-space distribution with \(r_q=2.0\).
This is predicted to be unstable by Figure \ref{fig:mw_l1_rq}.  The
modal power for the EXP simulation with \(N=1\times10^7\) particles
and the same distribution function is shown in Figure
\ref{fig:power_2p0}.  The growth for the most unstable value of
\(r_q\) is smaller than for the truncated models (cf. Figures
\ref{fig:mw_l1_trunc} and \ref{fig:mw_l1_rq}).  Nonetheless, the
predicted initial growth rates of the two modes are with 30\% of the
measured values, suggesting that the linear theory captures the
essential dynamics at early times.  Similar to the initially isotropic
case, the pattern speed of the eventual saturated \(l=1\) is slower
than for the linear mode.  The mSSA analysis of the simulation
demonstrates that the growing mode has the same character and
behaviour as the \(l=1\) instability in previous sections.

\subsection{Single power-law models}

\begin{figure}
  \includegraphics[width=0.45\textwidth]{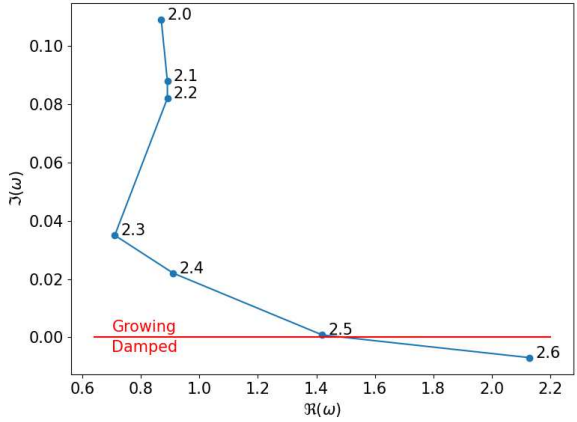}
  \caption{The location of zeros the \(\Re(\omega)\)--\(\Im(\omega)\)
    plane for a one-power model based on eq. (\ref{eq:NFW}) with
    \(\alpha\) indicated and \(\beta=0\) and an isotropic distribution
    function \(f(E)\) with truncation width 0.1 in virial units.  
    \label{fig:pow1_l1}}
\end{figure}

The failure to find an \(l=1\) unstable King model with an isotropic
distribution function (consistent with \citealt{Weinberg:91c})
motivated the consideration of single power-law models with tidal
truncations of equation (\ref{eq:two-power}) form with \(\beta=0\).
The mode diagram showing the zeros of the dispersion relation as a
function of the power-law exponent \(\alpha\) is shown in Figure
\ref{fig:pow1_l1}.  Power-law exponents \(\alpha\lesssim2.5\) are
\(l=1\) \emph{unstable} and and vice versa.  This suggests that an
extreme core-collapsed globular cluster could be susceptible.

For example, a classical core-collapsed power-law profile of
\(\alpha\approx2.3\) \citep[e.g.][]{Baumgardt.etal:2003} would have a
growth time of \(\sim 0.04\) in virial units.  A 100 pc globular
cluster with \(10^5\msun\) has a virial time of about 50 Myr,
suggesting an \(l=1\) growth time of 1 Gyr. In the simulations here,
the mode tends to saturate at roughly 10\% amplitude; this would
produce slight lopsided asymmetry as a function of isophotal radius.
There is a preferred axis, so observational detectablity depends on
line-of-sight orientation.  In addition, the ensemble of eccentric
orbits produced by the saturated mode may have implications for the
evolution of collisional systems.  For example, this population might
enhance core-halo momentum transfer and influence predictions made
under the assumption of spherically symmetric loss cones.

\section{Discussion and summary}
\label{sec:summary}

We explored the response of a collisionless halo to an \(l=1\)
(\emph{dipole}) distortion using two complementary techniques.
First, we performed N-body simulations and quantified the spatial
behaviour using a basis-function expansion (BFE).  We then used
multivariate singular spectrum analysis (mSSA) to identify the
correlated temporal changes in the spatial structure encoded by the
BFE.  For the N-body simulation, the initial distortions are imposed
by Poisson fluctuations.  Secondly, we used the linear response
operator derived for the same BFE to understand the resulting response
modes.  These modes are not the stationary modes, or by analogy with
the plasma literature, the \citet{Kampen:55} modes of the system, but
rather the phase-space structures that result from the linear
solution to an initial value problem.  For the linear theory, the
distortion is any function than can be represented by the expansion.
Therefore, the basis must be chosen carefully. The restriction to a
finite set of basis functions smooths the distribution to a chosen
scale. Conversely, a basis that over smooths important spatial scales
will lack sensitivity to small-scale modes. For the \(l=1\) mode
studied here, the mode is spatially large and easily represented by a
truncated basis function expansion.

As described in Section \ref{sec:intro}, the response modes are of
particular interest because they sustain themselves by self gravity
and can affect or \emph{dress} the response of our system at any
imposed nearby real frequency.  For stable systems, the frequency of
these modes have a negative imaginary parts (damped).  Most work on
stellar spheres considered stability, the most well-known results
being Antonov's stability criteria.  However, \citet{Weinberg:1994}
showed that the \(l=1\) modes are often weakly damped albeit stable.
Therefore, these may be excited and entrained by minor mergers in a
typical galaxian environment and noise in a quiescent environment.
Indeed, we have noticed in prior work that the \(l=1\) power in the
halo component is enhanced relative to \(l\ge2\) in \EXP\ simulations
and attributed this to dressed noise as described in
\citet{Weinberg:1994} and more recently by
\citet{Hamilton.Heinemann:2020}.  Tracking down the details of this
apparently excess power led to the identification of the instability
described in this paper.

The key results of this study are as follows:
\begin{enumerate}
\item The typically used idealised dark-halo models such as the NFW
  profile \citep{Navarro.Frenk.ea:97} are Antonov stable.  However, a
  minor modification that introduces another characteristic scale is
  enough to promote an \(l=1\) instability.  For an isotropic distribution
  function, a truncation between one and two virial radii will promote
  the instability.  The truncation induces an inflection in the phase
  space distribution function (i.e. some range of \(E\) such that
  \(df/dE>0\)) that powers the instability.
\item A similarly truncated \citet{Hernquist:90} model remains \(l=1\)
  stable.  A thorough investigation of outer power-law slopes of the
  NFW-like two-power family reveals that the onset of instability
  occurs for \(\beta\lesssim2.5\) in equation (\ref{eq:two-power}).
  This critical model is in between the outer power-law profiles of
  the NFW and Hernquist profiles.  This leads to a more general
  implication: the overall shape of the halo profile affects the
  long-term evolution of galaxies.
\item Single power-law halo profiles, also truncated at between one
  and two virial radii are unstable for \(r^{-\alpha}\) with
  \(\alpha\lesssim2.5\).  This may have implication for star clusters.
\item Truncation per se is not the important feature that determines
  growth.  Rather, a density band or ripple in the outer halo induces
  an inflection or bump in the phase space distribution function that
  can power the \(l=1\) growth.  We demonstrated this using
  Osipkov-Merritt distributions functions which are also unstable for
  modest values of anisotropy radius.  Their distribution functions
  have a similar \emph{bump} or inflection at high energies for finite
  anisotropy radius.
\item The exponentially growing mode transfers energy and angular
  momentum from the inner halo to the outer halo.  The affected orbits
  become more eccentric and bunched in radial angle as the mode grows.
  For our \(c=10\) NFW-like model, the peak \(l=1\) disturbance occurs
  at approximately 4 disc scale lengths.  Therefore, this mode is
  likely to influence the disc.
\item Section \ref{sec:instability} demonstrates that full realisation
  of a saturated \(l=1\) instability in a DM halo will require more
  time than the age of current Universe, \(t_0\).  However, scaled to
  the Milky Way, \(t_o\) corresponds to \(T=7\) in Figures
  \ref{fig:full} and \ref{fig:power_2p0}, and the amplitude is already
  significant.  The \(l=1\) mode in the DM halo is likely to affect
  the disc in the current epoch, even before the end of the
  exponential growth phase.  Moreover, the continued accretion of DM
  through cosmic time produces structure in the outer halo that may
  enhance coupling and promote instability.
\item The \(l=1\) mode is likely to be important in other systems
  with shorter time scales such as nuclear star clusters (without
  SMBHs) and possibly globular clusters. The saturated mode presents
  as a comoving, in phase, overdensity of orbits which may have
  consequence for collision rates and momentum exchange.
\end{enumerate}

\section*{Acknowledgements}
MDW would like to thank the SEGAL collaboration and J.-B. Fouvry in
particular who restimulated my interest in this problem and the
referee whose questions and attention detail stimulated some new
findings and improved the presentation.  MDW also thanks Mike Petersen
for his many comments and suggestions.

\section*{Data availability}

The data underlying this article will be shared on reasonable request
to the author.

\bibliographystyle{mn2e}

\appendix

\section{The EXP implementation of the biorthogonal
  expansion}
\label{sec:biorth}

\EXP\ allows for a straightforward calculation of the gravitational
potential from the mass distribution through time.  The key limitation
of the BFE method lies in the loss of flexibility owing to the
truncation of the expansion: large deviations from the equilibrium
disc or halo will not be well represented. Although the basis is
formally complete, a truncated expansion limits the variations that
can be accurately reconstructed. Despite this, basis functions can be
a powerful tool for physical insight; analogous to traditional Fourier
analysis, a BFE identifies spatial scales and locations responsible
for dynamical evolution.  In particular, this paper investigates
relatively low-amplitude, large-scale distortions that are well
represented by the expansion.  Our spherical expansions truncate the
series by spherical harmonic order, \(l_{max}\), and radial order
\(n_{max}\).

\EXP\ optimally represents the BFE for haloes with radial basis
functions determined by the target density profile and spherical
harmonics. The lowest-order \(l=0, m=0\) basis function matches the
potential and density of the equilibrium.  Each successive basis
function at each harmonic order \(l\) introduces variation at a
successively smaller spatial scale.  Therefore, truncation of this
basis function series limits the spatial sensitivity to some minimum
spatial scale.  For an N-body simulation, this truncation effectively
removes the small-scale \(1/N\) fluctuations.  Therefore, the BFE
method would not be an appropriate technique for studying the details
of two-body relaxation.  Conversely, this is an advantage for
approximating a collisionless system.

\subsection{Method summary}

Any disturbance in an approximately spherical system may be
represented as a spherical harmonic expansion with an appropriate set
of orthogonal radial wave functions.  We choose the biorthogonal
potential-density pairs described in \citet{Petersen.etal:2022}.  The
pair of functions \((u^{lm}_i,d^{lm}_i)\) is constructed to satisfy
Poisson's equation, \(\nabla^2u^{lm}_i = 4\pi Gd^{lm}_i\), and to form
a complete set of functions with the scalar the product
\begin{equation}
	-\frac{1}{4\pi G}\int dr\,r^2 u^{lm\,\ast}_i(r) d^{lm}_j(r) =
	\begin{cases}1& \mbox{if}\ i=j \\ 0 & \mbox{otherwise.}\end{cases}
\end{equation}
The total density and potential then have the following expansions:
\begin{eqnarray}
  \tilde{\Phi}(\mathbf{r}) &=& \sum_{lm}Y_{lm}(\theta,\phi)\sum_j a^{lm}_j(t)
  u^{lm}_j(r),		\label{eq:exp1}	\\
  \tilde{\rho}(\mathbf{r}) &=& \sum_{lm}Y_{lm}(\theta,\phi)\sum_j a^{lm}_j(t)
  d^{lm}_j(r).		\label{eq:exp2}
\end{eqnarray}
The gravitational potential energy in these fields are:
\begin{equation}
  W=\frac12\int d^3r
  \tilde{\rho}(\mathbf{r})\tilde{\Phi}(\mathbf{r}) = -\frac12\sum_{lmj}
  |a^{lm}_j|^2,
  \label{eq:power}
\end{equation}
having used the orthogonality properties of the spherical harmonics
and biorthogonal functions.  The partial sums of the squared
coefficients of the expansion are then twice the gravitational
potential energy in each harmonic order \(l\).

When applied to an N-body simulation, we obtain time series
\(\mathbf{a}^{lm}_j\) for each coefficient in sums in equations
(\ref{eq:exp1}) and (\ref{eq:exp2}).  See \citet{Petersen.etal:2022}
for an in-depth discussion of N-body simulation with this
gravitational field estimator.

\subsection{Parameter choices}

The overall fields of the halo are described by
\(\left(l_{\rm halo}+1\right)^2\times n_{\rm halo}\) terms, where
\(l_{\rm halo}\) is the maximum order of spherical harmonics and
\(n_{\rm halo}\) is the maximum order of radial terms per \(l\) order.
For simulations and analyses in this paper, we use a maximum harmonic
order \(l_{\mbox{\tiny max}}=6\) and maximum radial order
\(n_{\mbox{\tiny max}}=20\) using the Sturm-Liouville basis
conditioned on each particular input model.  While the quality of the
expansion is centre dependent, the basis could follow any displacement
from the centre for sufficiently large particle number and large
values of \(l_{\mbox{\tiny max}}\) and \(n_{\mbox{\tiny max}}\).  The
choices of \(l_{\mbox{\tiny max}}=6\) and \(n_{\mbox{\tiny max}}=20\)
allows us to follow centre displacements seen in our simulations (see
Secs. \ref{sec:center} and \ref{sec:verify} for more discussion).
These series of coefficients efficiently describe time dependence of
the gravitational field produced by the N-body evolution\footnote{This
  method will work for triaxial halos as well.  For triaxial halos,
  the target density can be chosen to be a close fitting spherical
  approximation of the triaxial model.  The equilibrium will require
  non-axisymmetric terms but the series will converge quickly.}.

\subsection{Centring and momentum conservation}
\label{sec:center}

BFE codes such as \EXP\ require a choice for the expansion centre.
\EXP\ monitors the centre of potential by sorting the particles in
gravitational energy and computing the centre-of-mass location from
the ball enclosing the most bound particles.  We have verified that
the expansion can reproduce centre shifts of our simulations by
showing that two simulations that use the original centre and the
centre determined by the most bound particles reproduce the same
overall evolution.

Moreover, our research topic is \(l=1\) modes in an empty environment.
Therefore, dynamical consistency demands that the linear momentum of
the entire system be conserved in the presence of a \(l=1\) mode.
Indeed, this conservation is nearly reproduced by \EXP; the computed
centre of mass for these simulations remains on the initial centre to
approximately 0.001 virial radii.  The BFE approach does not
explicitly conserve momentum by construction, so this is a strong
consistency check.  Nonetheless, to put to rest concerns that results
are affected by centring artefacts, we repeated our fiducial
simulation using a tree-gravity Poisson solver which is centre
independent by construction; see Section \ref{sec:verify}.

\section{The biorthogonal dispersion relation}
\label{sec:disper}

\subsection{The response matrix}

We follow the procedure described in \citet{Weinberg:1994} to reduce
the coupled system of the collisionless Boltzmann and Poisson
equations to a solution of a matrix equation using the expansion
above.  The solution may then be written as the non-trivial solution
to the following linear equation
\begin{equation}
  {\hat a}^{lm}_i = M_{ij}(\omega){\hat a}^{lm}_j,
        \label{eq:sphresp}
\end{equation}
where summations over like indices are implied, the `\(\hat{\ }\)'
indicates a Laplace-transformed quantity, \(\omega\) is a
complex-valued frequency and
\begin{eqnarray}
	M_{ij}(\omega)
	&=& \frac{(2\pi)^3}{4\pi G}\int\int \frac{dE\,dJ\, J}{\Omega_1(E,J)}
	\sum_{{\bf n}}\frac{2}{2l+1} \times\nonumber \\
            &&{\bf n}\cdot\frac{\partial f_0}{\partial{\bf I}}
        \frac{1}{\omega-{\bf n}\cdot{\bf\Omega}}
        \left|Y_{l,l_2}(\pi/2,0)\right|^2 \times \nonumber \\
            && W^{l_1,\,i\ \ast}_{l,l_2,l_3}({\bf I})
        W^{l_1,\,j}_{l,l_2,l_3}({\bf I}),
        \label{eq:sphmat}
\end{eqnarray}
where
\begin{equation}
  W^{l_1}_{l\,l_2l_3}(I_1,I_2) = \frac{1}{\pi}\int^\pi_0 dw_1\cos[l_1w_1-l_2
              (\psi-w_2)]u_{j}^{l\,l_3}(r).
  \label{eq:potrans2}
\end{equation}
In the above equations, \(I_1, I_2, I_3\) are the actions,
\(w_1, w_2, w_3\) are the angles corresponding to the radial,
tangential, and azimuthal action-angle variables and
\({\bf n}=(l_1,l_2,l_3)\) is a vector of integers.  The actions were
chosen by solution of the Hamiltonian-Jacobi equation; \(I_1\) is the
radial action, \(I_2\) is the total angular momentum and \(I_3\) is
the z-projection of the angular momentum.  The quantity \(\psi-w_2\),
the difference between the position angle of a star in its orbital
plane and its mean azimuthal angle, depends only on \(w_1\), the angle
describing the radial phase.  The frequencies associated with the
conjugate angles \({\bf w}\) are defined by
\({\bf\Omega}=\partial H/\partial{\bf I}\).  The quantity \(\Omega_1\)
is the radial frequency and reduces to the usual epicyclic frequency
for nearly circular orbits.  The quantity \(\Omega_2\) is the mean
tangential frequency and reduces to the circular orbit frequency,
\(v_c(r)/r\), for orbits with circular velocity \(v_c\).  The quantity
\(\Omega_3=0\) and the corresponding angle \(w_3\) is the azimuth of
the ascending node (see Tremaine and Weinberg
1984\nocite{Tremaine.Weinberg:84}, for details).  It follows that
\(M_{ij}^\ast(\omega)=M_{ij}(-\omega^\ast)\) and this property is
exploited in the computations described in \S3 to reduce the number of
numerical evaluations.  Since most researchers quote phase-space
distribution functions in energy and total angular momentum
\(f(E,J)\), We have transformed the integration in the definition of
\({\bf M}(\omega)\) in equation (\ref{eq:sphmat}) to \(E\) and \(J\).

\subsection{The dispersion relation}

Equation (\ref{eq:sphresp}) only has a nontrivial solution if
\begin{equation}
{\cal D}(\omega)\equiv\hbox{det}\{{\bf 1}-{\bf M}(\omega)\}=0.
\label{eq:sphdisp}
\end{equation}
As in plasma theory, we refer to the function \({\cal D}(\omega)\) as
a dispersion relation.  In general, the value of \(\omega\) solving
equation (\ref{eq:sphdisp}) will be complex.  The coefficients
\(\hat{a}_j\) which appear in equation (\ref{eq:sphresp}) are Laplace
transformed and describe the response of the system to a perturbation
with a time-dependence of the form \(\exp(-i\omega t)\).  Because
equation (\ref{eq:sphresp}) follows from a Laplace transform, it is
only valid in its present form for \(\omega\) in the upper--half
complex plane.  The dispersion relation must be analytically continued
to \(\Im(\omega)<0\) to find damped modes.  The response coefficients
\(\hat{\bf a}^{lm}\) which solve equation (\ref{eq:sphresp}) for a
particular eigenfrequency \(\omega\), describes a response mode of the
system.

Because a sphere has no unique symmetry axis, a response mode can not
depend on an arbitrary choice of coordinate axes.  Since a rotation
causes the azimuthal components indexed by \(m\) to mix according to
the rotation matrices, the mode itself must depend on \(l\) alone.  In
particular, the functions \((u^{lm}_i,d^{lm}_i)\) may be chosen
independent of \(m\) (e.g. the spherical Bessel function used by
Fridman and Polyachenko 1984, and Weinberg
1989\nocite{Fridman.Polyachenko:84b,Weinberg:89}).  Therefore,
equation (\ref{eq:sphmat}) and the dispersion relation are independent
of \(m\).

\subsection{Convergence}

The dispersion relation depends on the response matrix
(eq. \ref{eq:sphmat}).  The elements of this matrix require a
two-dimensional quadrature in actions and the computation of the
action-angle transform of the biorthogonal potential functions
(eq. \ref{eq:potrans2}).

The integral in equation (\ref{eq:potrans2}) defining the angle-action
transform is a periodic and therefore has a converging trigonometric
expansion owing to the smoothness of the finite-order biorthogonal
functions. The Euler-Maclaurin summation formula guarantees that
evaluation of a periodic integrand on a uniform grid with \(k\) knots
will have a negligible error term for sufficiently large \(k\).  Since
\(n_{max}=20\), there are at most 20 oscillations in the biorthogonal
potential functions.  Empirically, a choice of \(k=200\) is more than
sufficient to compute the \(W^{l_1}_{l\,l_2l_3}(I_1,I_2)\) to 1 part
in \(10^5\).  With these choices, the error in evaluating equation
(\ref{eq:potrans2}) is much smaller than the error in the double
quadrature in equation (\ref{eq:sphmat}).

Turning to the integral in equation (\ref{eq:sphmat}), the resonant
denominator in the integrand is the main source of numerical
difficulty.  Rather than use the actions \(I_1, I_2\) as quadrature
variables, we transformed from \(I_1, I_2\) to energy and scaled
angular momentum variables \(E, \kappa\) where
\(\kappa \equiv I_2/J_{max}(E)\) and \(J_{max}(E)\) is the angular
momentum of the circular orbit with energy \(E\).  This change was
motivated by the behaviour of the
\({\bf n}\cdot{\bf\Omega}\approx\mbox{constant}\) loci in each
coordinate system.  The loci are anti-diagonal in actions
\((I_1, I_2)\) but tend to lie along lines of constant \(\kappa\) in
the new variables.  Therefore, the choice of \((E, \kappa)\) reduces
the total number of necessary quadrature knots.  We choose
\(2^{(n_E)}\) and \(2^{(n_\kappa)}\) knots in \(E\) and \(\kappa\),
respectively.

The matrix elements in equation (\ref{eq:sphmat}) also depend on a
sum over the integers \({\bf n}\).  For the \(l=1\) harmonic
considered here, \(l_1\in(-\infty, \ldots, \infty)\),
\(l_2\in[-1, 1]\), and \(l_3\in[-1, 1]\).  We truncate the sum in
\(l_1\) with the choice \(l_1\in[-l_{1,max}, l_{1, max}]\).  In
practice, the sum converges quickly in \(l_{1, max}\) for two reasons:
(1) only a small combination of \({\bf n}\) lead to resonant
denominators; and (2) the potential transform (eq. \ref{eq:potrans2})
is a Fourier integral which decreases exponentially as
\(|l_1|\rightarrow\infty\).

\begin{figure}
  \includegraphics[width=0.45\textwidth]{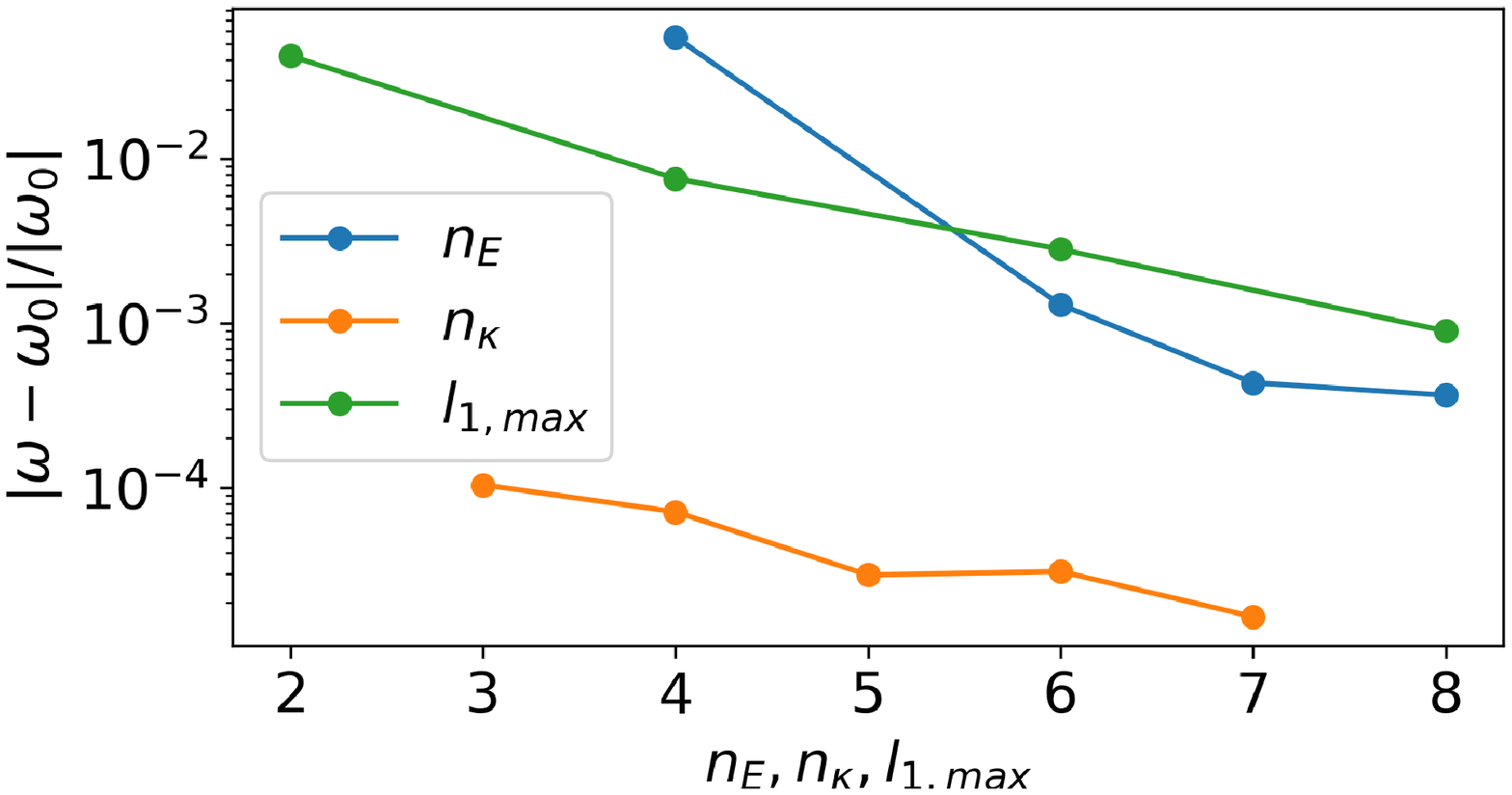}
  \caption{The relative difference for the zero of the dispersion,
    as each of the parameters \(n_E, n_\kappa, l_{1,max}\)
    are increased in turn.  Each solution, \(\omega\), is compared
    to the solution for the largest parameter value in the sequence,
    \(\omega_0\). The complex frequency of this mode is
    \(\omega_0 \approx (1.83, 0.11)\).  The two non-varying parameters
    in each sequence are set to \(n_E=7, n_\kappa=5, l_{1,max}=8\).
    The series converge with relative errors in \(\omega\) are \({\cal
      O}(10^{-3})\) for the working parameter set.
    \label{fig:converge}}
\end{figure}

We demonstrate convergence by exploring the approach to the zero of
the dispersion relation (eq. \ref{eq:sphdisp}) for the unstable
\(l=1\) mode described in Section \ref{sec:instability}.  The complex
frequency of this mode is \(\omega_0 \approx (1.83, 0.11)\).  First,
we choose a set of fiducial values for the parameters which
empirically \(n_E=7, n_\kappa=5, l_{1,max}=8\) which are close to
converged.  We evaluate sequences of \(\omega\) with
\({\cal D}(\omega)=0\) by increasing each of
\(n_E, n_\kappa, l_{1,max}\) in turn while the others remain at their
fiducial value.  Figure \ref{fig:converge} shows the modulus of the
difference between these parameters and the last one in the sequence.
The relative error in \(\omega_0\) decreases as each of
\(n_E, n_\kappa, l_{1,max}\) increase.  As expected, the dependence on
\(n_\kappa\) is weak while the convergence in \(n_E\) is slower,
approaching a relative error of approximately \(5\times10^{-4}\) af
the end of the sequence.  The larger values of \(l_1\) do affect the
location of \(\omega_0\) but \(l_{1,max}=8\) does seem sufficient for
the purposes of this work.

\subsection{Evaluating the response matrix}
\label{sec:cusp}

\subsubsection{Application to cuspy halo}

We evaluate a grid of orbital parameters at the positions of the
\(E, \kappa\) knots used to perform the quadratures for the the matrix
elements (eq. \ref{eq:sphmat}).  At each pair \(E, \kappa\), the
linear response code computes the actions \(I_1, I_2\), the inner and
outer turning points, the frequencies \(\Omega_1, \Omega_2\) and the
run of radius as a function of the radial angle and the difference
between the azimuth and mean azimuth, \(\psi - w_2\).  For the
\(r^{-1}\) cusps considered here (see Sec. \ref{sec:models}), the
orbital frequencies are proportional to \(r^{-1/2}\).  While there is
no fundamental numerical problem computing these values in a cusp,
including a core radius in the model results in bounded frequency
values and prevents hand-tuning the root finders used to compute
orbital turning points.  One can accomplish the same result by
choosing the lower end point of the energy integral offset very
slightly from the potential value at \(r=0\).

For the small values of \(r_c\) used for our fiducial model, the mass
fraction inside of \(r_c\) is approximately \(3\times10^{-7}\) of the
total.  It is possible that our choice of finite \(r_c\) could affect
the value of \({\cal D}(\omega)\) for values of
\(\omega\sim\Omega_{1,2}\) at radii inside of \(r_c\).  However, these
orbital frequencies are so large compared to our modal frequencies
that that resonant coupling is not possible.  On the N-body side, we
have repeated the simulation with zero core radius and there is no
distinguishable difference.

\subsubsection{Mode location}
\label{sec:modeloc}

The zeros of the dispersion relation, equation (\ref{eq:sphdisp}), that
determine the point-mode locations are found in two steps:
\begin{enumerate}
\item We begin by evaluating \({\cal D}(\omega)\) over a coarse grid
  with 20 or 40 real frequencies and 5 or 10 imaginary frequencies.
  We know a priori that the real frequencies of the weakly damped or
  weakly growing modes will be in the range \(\Re(\omega)=[0, 20]\) in
  units where the gravitational constant, mass and outer radius are
  unity.  We choose \(\Im(\omega)=[0, 0.2]\).  A visual inspection of
  the \(|{\cal D}(\omega)|\) surface will reveal the locations of an
  unstable mode in the upper-half plane as a zero or a damped mode in
  the lower-half as a trough decreasing towards
  \(\Im(\omega)\rightarrow0\).
\item If the mode is in the upper half plane, we attempt to isolate
  the mode with successively refined \(10\times10\) grids bracketing
  the zero in \(|{\cal D}(\omega)|\).  We determine the mode by using
  bilinear interpolation to find the zero-valued loci in
  \(\Re({\cal D})\), \(\Im({\cal D})\) and find the intersection of
  these two loci to estimate the solution.  If the mode is in the
  lower half plane, we use the rational function extrapolation
  technique described in \citet{Weinberg:1994} to estimate values of
  \({\cal D}(\omega)\) in the vicinity of the zero.  We recommend
  using 30 or fewer \(\omega\) points as input to the rational
  function construction algorithm.  For the results in Section
  \ref{sec:linear}, we typically used used a grid of 8 in
  \(\Re({\omega})\) by 3 in \(\Im({\omega})\) spaced around the trough
  in \(\Re({\omega})\) and a height in \(\Im({\omega})\) of order the
  scale length in the trough's gradient.  Large numbers of \(\omega\)
  points may result in nearly cancelling roots in the numerator and
  denominator of the rational function.  These lead to numerical
  artefacts from incomplete cancellation. It should be possible to
  apply a heuristic to remove these pairs of roots, but that has not
  been done here.
\end{enumerate}

\subsection{The \(l=1\) neutral translation mode}
\label{sec:trmode}

In addition to the point modes that we use interpret our simulations
described in Section \ref{sec:linear}, the response matrix admits a
zero-frequency \(l=1\) mode.  This point mode corresponds to a body
displacement of the system.  Conservation of linear momentum prohibits
self excitation of this mode; it can only be excited by external
perturbation that applies a net force to the initially unperturbed
system.  The conservation of linear momentum has been addressed in
Appendix \ref{sec:center}.

In addition, the biorthogonal basis derived using solutions of the
Sturm-Liouville equation \citep{Weinberg:99,Petersen.etal:2022}
provides variation in and around the characteristic radius of the
system but not at small and large radii.  Therefore, the translation
mode converges very slowly in the cusp.  Fortunately, this mode has no
bearing on the analyses in \S \ref{sec:linear}.

\subsection{Derivation of the point-mode shape and the system
  response}
\label{sec:modal_response}

The modal eigenvector \(\hat{a}_j\) can be easily found through
singular value analysis of the response matrix (eq. \ref{eq:sphmat})
evaluated at the modal frequency.  The desired eigenvector will have
eigenvalue \(\lambda\approx1\).  The proximity of \(\lambda\) to \(1\)
provides a self-consistency check on the numerics of the dispersion
relation.  Typically, these values are \(\lambda=1\pm0.05\).  The
density and potential perturbations follow directly using equations
(\ref{eq:exp1}) and (\ref{eq:exp2}).  For an unstable mode, it is
sufficient to determine the eigenvalues and eigenvectors of
\({\cal D}(\omega)\) at the modal frequency.  For a stable damped
mode, the matrix elements can be estimated by the same rational
function extrapolation method used in Section \ref{sec:modeloc} (see
\citealt{Weinberg:94}).

\label{lastpage}

\end{document}